\documentclass[12pt]{SelfArx} 
\usepackage[english]{babel} 
\usepackage{amsfonts,amssymb,amsbsy,amsmath,amsthm}
\usepackage{listings}
\usepackage{lscape}
\usepackage{multirow}
\usepackage{threeparttable}
\usepackage{here}
\usepackage{url}
\usepackage[numbers]{natbib}

\theoremstyle{definition}

\theoremstyle{plain}

\definecolor{color1}{RGB}{0,0,90} 
\definecolor{color2}{RGB}{0,20,20} 

\usepackage{hyperref} 
\hypersetup{hidelinks,colorlinks,breaklinks=true,urlcolor=red,citecolor=color1,linkcolor=color1,bookmarksopen=false,
	pdftitle={Sample size calculations for single-arm survival studies using transformations of the Kaplan--Meier estimator},
	pdfauthor={Kengo Nagashima, Hisashi Noma, Yasunori Sato, Masahiko Gosho}}

\JournalInfo{\emph{Pharmaceutical Statistics} 2021; \textbf{20}(3): 499--511.} 
\Archive{DOI: 10.1002/pst.2090} 

\PaperTitle{Sample size calculations for single-arm survival studies using transformations of the Kaplan--Meier estimator}
\ShortTitle{Sample size calculations for the Kaplan--Meier estimator}

\Authors{Kengo Nagashima\textsuperscript{1}*, Hisashi Noma\textsuperscript{2}, Yasunori Sato\textsuperscript{3}, and Masahiko Gosho\textsuperscript{4}} 
\affiliation{\footnotesize{\textsuperscript{1}\textit{Research Center for Medical and Health Data Science, The Institute of Statistical Mathematics, Japan.}}}
\affiliation{\footnotesize{\textsuperscript{2}\textit{Department of Data Science, The Institute of Statistical Mathematics, Japan.}}}
\affiliation{\footnotesize{\textsuperscript{3}\textit{Department of Preventive Medicine and Public Health, Keio University School of Medicine, Tokyo, Japan.}}}
\affiliation{\footnotesize{\textsuperscript{4}\textit{Department of Biostatistics, Faculty of Medicine, University of Tsukuba, Ibaraki, Japan.}}}
\affiliation{\footnotesize{*\textbf{Corresponding author}: nshi1201@gmail.com}}
\affiliation{}
\affiliation{This is a preprint/accepted version (27-Oct-2020) of the following manuscript: Nagashima K, Noma H, Sato Y, Gosho M. Sample size calculations for single-arm survival studies using transformations of the Kaplan--Meier estimator. \emph{Pharmaceutical Statistics} 2021; \textbf{20}(3): 499--511.}

\Keywords{
\footnotesize{
Survival analysis, Sample size, Kaplan--Meier estimator, Confidence intervals, One sample test
}
} 
\newcommand{\keywordname}{Keywords} 

\Abstract{
\footnotesize{
In single-arm clinical trials with survival outcomes, the Kaplan--Meier estimator and its confidence interval are widely used to assess survival probability and median survival time.
Since the asymptotic normality of the Kaplan--Meier estimator is a common result, the sample size calculation methods have not been studied in depth.
An existing sample size calculation method is founded on the asymptotic normality of the Kaplan--Meier estimator using the log transformation.
However, the small sample properties of the log transformed estimator are quite poor in small sample sizes (which are typical situations in single-arm trials), and the existing method uses an inappropriate standard normal approximation to calculate sample sizes.
These issues can seriously influence the accuracy of results. In this paper, we propose alternative methods to determine sample sizes based on a valid standard normal approximation with several transformations that may give an accurate normal approximation even with small sample sizes.
In numerical evaluations via simulations, some of the proposed methods provided more accurate results, and the empirical power of the proposed method with the arcsine square-root transformation tended to be closer to a prescribed power than the other transformations.
These results were supported when methods were applied to data from three clinical trials.
}
}

\begin{document}

\flushbottom 

\maketitle 


\thispagestyle{empty} 


\section{Introduction}
\label{sec:1}

In single-arm clinical trials with survival outcomes, the Kaplan--Meier estimator and its pointwise confidence interval are widely used to assess survival probability at a specific time, as well as median survival time.
For instance, single-arm survival designs have been used in early phase studies for metastatic colorectal cancer using molecular-targeted drugs \citep{Sidhu2013} and advanced hepatocellular carcinoma \citep{Llovet2019}.
Progression-free survival is known as a validated surrogate endpoint for metastatic colorectal cancer, and progression-free survival rate at a specified time is a frequently used endpoint in single-arm studies.
For advanced hepatocellular carcinoma, sorafenib was shown to have clear survival benefits, but it was only associated with tumor response of 2--3\%.
Overall survival or progression-free survival rates are frequently used as the primary endpoint in single-arm studies of advanced hepatocellular carcinoma.
These designs have been used in other oncology studies.
Single-arm survival designs have also been used in studies could not include placebo administration, such as clinical trials for childhood and rare diseases \citep{Slayton2018}.
These studies are usually conducted with small sample sizes.

The asymptotic normality of the Kaplan--Meier estimator is well-known \citep{Breslow1974}, and approximation is quite poor with small sample sizes \citep{Borgan1990}.
To resolve the issues of small sample size when inferring the Kaplan--Meier estimator, several alternative methods have been suggested to construct confidence intervals and hypothesis tests using transformations, such as the log transformation \citep{Borgan1990}, the log-minus-log transformation \citep{Kalbfleisch1980}, the logit transformation \citep{Meeker1998}, and the arcsine square-root transformation \citep{Thomas1975}.
It is important to choose a suitable transformed statistic when calculating sample size in single-arm clinical trials with survival outcomes because small sample sizes are common.

Since the asymptotic property of the Kaplan--Meier estimator has been extensively discussed, sample size methods have not been studied in depth.
The commonly used standard sample size calculation method by Cancer Research And Biostatistics \cite{CRAB2002} also uses an asymptotic normality assumption of the Kaplan--Meier estimator with the log transformation.
However, the performance of the log transformed estimator is quite poor under the typical small sample conditions of these single-arm trials \citep{Borgan1990}, and the existing method uses an inappropriate standard normal approximation to calculate sample sizes to compensate for the poor performance.
These problems can influence the accuracy of results from sample size calculations.
An alternative approach that is based on the exact method has been proposed for the small sample problem \cite{Chang1996}.
The exact method guarantees the type I error rate.
However, the method is commonly conservative when sample size is small.
In addition, the method would be computationally intractable when sample size is moderate ($n > 20$) because the method requires computing exact permutation distributions.
Several papers discussed about related topics (e.g., Wu and Xiong \cite{Wu2014} discussed about sample size methods about the Nelson--Aalen estimator and Wu \cite{Wu2015} proposed alternative method based on the one-sample log-rank test).
However, the Kaplan--Meier estimator and its pointwise confidence interval are widely used in practice.
Thus, we focus on sample size methods for the Kaplan--Meier estimator based on transformations.

In this article, we consider alternative sample size calculation methods for the Kaplan--Meier estimator that use the transformations mentioned above and an appropriate standard normal approximation, and we discuss their theoretical properties in Section 2.
In Section 3, we assess the performance of the existing and proposed methods in simulation studies because the small sample properties of hypothesis tests based on the Kaplan--Meier estimator and sample size methods highly depend on the type of transformations.
We also demonstrate the effectiveness of the proposed method via simulation based on three clinical trials in Section 4.
The goal of this paper is to give recommendations to improve sample size calculation methods for the Kaplan--Meier estimator in single-arm survival studies.

\section{Sample size calculation methods for the Kaplan--Meier estimator}
\label{sec:2}

\subsection{Asymptotic results of the Kaplan--Meier estimator}
\label{sec:2.1}
We considered asymptotic distributions of the Kaplan--Meier estimator \citep{Kaplan1958}, $\hat{S}(t)$, and its transformation, $g\{\hat{S}(t)\}$, with a function $g(\cdot)$.
Let $T_i$ for $i=1,\ldots,n$ be a random variable with a survival time that is distributed as a distribution function $F(t)$, $U_i$ be a random variable with a censoring time that is distributed as a specific distribution function, $X_i=\min\{T_i,U_i\}$ be the observed time, $\delta_i=I(T_i < U_i)$ be a censoring indicator, and $I(\cdot)$ be an indicator function with a value of $1$ if its argument is true, and zero otherwise.
At time $t$, the Kaplan--Meier estimator is defined as,
\[
\hat{S}(t)=
\prod_{s<t}\left\{
1-\frac{\Delta\bar{N}(s)}{\bar{Y}(s)}
\right\},
\]
and its asymptotic variance estimator is defined as,
\[
\hat{\sigma}^2(t)=
\hat{S}(t)^2 \int_{0}^{t} \frac{1}{\bar{Y}(s)\{\bar{Y}(s)-\Delta \bar{N}(s)\}} \mathrm{d}\bar{N}(s),
\]
where $\bar{N}(t)=\sum_{i=1}^{n}N_i(t)$, $\bar{Y}(t)=\sum_{i=1}^{n}Y_i(t)$, $\Delta X(t)=X(t)-X(t-)$, $X(t-)=\lim_{h\downarrow 0}X(t-h)$, $N_i(t)=I(X_i \leq t, \delta_i=1)$ is a counting process that counts the number of failures ($0$ or $1$) for the $i$-th individual, and $Y_i(t)=I(X_i \geq t)$ is an at risk process that takes the value of $1$ if the $i$-th individual is at risk at time $t$ and $0$ otherwise.
Let $\Lambda(t)=-\log S(t)=\int_0^t \lambda(s) \mathrm{d}s$ be a cumulative hazard function, $\lambda(t)=-\frac{1}{S(t)}\frac{\mathrm{d}S(t)}{\mathrm{d}t}$ be a hazard function, and $\hat{\Lambda}(t)=\int_0^t \{\bar{Y}(t)\}^{-1} \mathrm{d}\bar{N}(s)$ be the Nelson--Aalen estimator \citep{Nelson1969,Aalen1978}.
By the martingale central limit theorem, $L_n(t)=\sqrt{n}\{\hat{\Lambda}(t)-\Lambda(t)\}=\int_{0}^{t} \sqrt{n}\{\bar{Y}(s)\}^{-1} \mathrm{d}\bar{M}(t) \to_L L(t)$, where $\bar{M}(t)=\bar{N}(t)-\int_0^t \bar{Y}(s)\lambda(s) \mathrm{d}s$, $L(t)$ is a Brownian motion with $\mathrm{E}[L(t)]=0$ and $\mathrm{Var}[L(t)]=\lim_{n\to\infty} \langle L_n, L_n \rangle(t)=\int_0^t \{\mathrm{P}(U>s)S(s)\}^{-1} \mathrm{d}\Lambda(s)$ \citep{Fleming1991}.
The Kaplan--Meier and Nelson--Aalen estimators have the relationship $\sqrt{n}\{\hat{S}(t)-S(t)\}=\sqrt{n}[\exp\{-\hat{\Lambda}(t)\}-\exp\{-\Lambda(t)\}]+o_P(1)$ \citep{Breslow1974}.
By applying the functional delta method, the asymptotic distribution of the Kaplan--Meier estimator can be obtained as $\sqrt{n}\{\hat{S}(t)-S(t)\} \to_L -S(t)L(t)$ \citep{Gill1990}, and the asymptotic variance of the Kaplan--Meier estimator is $\sigma^2(t)=\mathrm{Var}[\sqrt{n}\{\hat{S}(t)-S(t)\}]=S^2(t)\int_0^t \{\mathrm{P}(U>s)S(s)\}^{-1} \mathrm{d}\Lambda(s)$ \citep{Andersen1993}.
Under small sample sizes, the accuracy of the normal approximation of the Kaplan--Meier estimator is not sufficient, especially when the value of the survival function is close to $0$ or $1$ \citep{Borgan1990}.
Several transformations have been used to improve its accuracy, which are defined as follows:
\begin{enumerate}
\item Identity transformation \citep{Kaplan1958} (corresponds to an estimator for $S(t)$): $g\{S(t)\}=S(t)$.
\item log transformation \citep{Borgan1990} (corresponds to an estimator for $\Lambda(t)$): $g\{S(t)\}=\log S(t)$.
\item log-minus-log transformation \citep{Kalbfleisch1980}: $g\{S(t)\}=\log\{-\log S(t)\}$.
\item logit transformation \citep{Meeker1998}: $g\{S(t)\}=\log[S(t)\{1-S(t)\}^{-1}]$.
\item arcsine square-root transformation \citep{Thomas1975}: $g\{S(t)\}=\arcsin \sqrt{S(t)}$.
\end{enumerate}
Several studies recommend the log-minus-log \citep{Kalbfleisch1980}, arcsine- \citep{Thomas1975}, or both \citep{Borgan1990,Aalen2008}, transformed confidence intervals of the Kaplan--Meier estimator, since these have performed better than the identity and log transformations.
Borgan and List{\o}l \cite{Borgan1990} discussed why the transformed confidence intervals have better small sample properties, finding that the transformed confidence intervals have a less skewed distribution than non-transformed confidence intervals.
Borgan and List{\o}l \cite{Borgan1990} also discussed the asymptotic normality of transformations of the Kaplan--Meier estimator.
This result is summarized as follows: Suppose that a transformation $g$ has the derivative $g'$ at the point $S(t)$ and $g'$ is not zero, for $0 \leq t \leq T$; Then
\begin{equation}
\label{equ:1}
\sqrt{n}[g\{\hat{S}(t)\} - g\{S(t)\}] \to_L g'\{S(t)\}\{-S(t)L(t)\},
\end{equation}
and $\tau^2(t)=\mathrm{Var}(\sqrt{n}[g\{\hat{S}(t)\} - g\{S(t)\}])=g'\{S(t)\}^2 \sigma^2(t)$.

For each transformation, the derivative $g'\{S(t)\}=\frac{\mathrm{d} g\{S(t)\}}{\mathrm{d}S(t)}$ is as follows:
\begin{enumerate}
\item identity transformation: $g'\{S(t)\}=1$.
\item log transformation: $g'\{S(t)\}=\{S(t)\}^{-1}$.
\item log-minus-log transformation: $g'\{S(t)\}=\{S(t)\log S(t)\}^{-1}$.
\item logit transformation: $g'\{S(t)\}=[S(t) \{1-S(t)\}]^{-1}$.
\item arcsine square-root transformation: $g'\{S(t)\}=[4S(t) \{1-S(t)\}]^{-1/2}$.
\end{enumerate}
The variance  $\tau^2(t)=g'\{S(t)\}^2 \sigma^2(t)$ can be obtained by replacing $g'\{S(t)\}^2$ by the appropriate term above.
It should be noted that the assumptions of the derivative will hold if $0 < S(t) \leq 1$ for the log transformation, or $0 < S(t) < 1$ for the log-minus-log, logit, and arcsine square-root transformations.

\subsection{Proposed methods to determine sample size}
\label{sec:2.3}

We apply a straightforward sample size formula.
Here, we consider one-sided hypotheses, $H_0: \epsilon \leq 0$, and $H_1: \epsilon > 0$, where $\epsilon = g\{S_1(t)\} - g\{S_0(t)\}$ is a transformed effect size, and $S_0(t)$ and $S_1(t)$ are survival functions at time $t$ under the null and alternative hypotheses.
The analysis time $t$ is a specific time point and must be determined at the design stage.
In general, time $t$ is defined from a clinical and/or statistical perspective.
Annual survival probabilities (e.g., 6-month progression free survival, 1-year overall survival, and so on) are widely used in practice by taking into account clinical importance and high statistical power.
We reject the null hypothesis at the type-I error rate $\alpha$, if
\[
Z=\frac{g\{\hat{S}(t)\} - g\{S_0(t)\}}{\hat{\tau}(t)/\sqrt{n}}>z_{1-\alpha},
\]
based on $Z \to_L N(0, 1)$ as $n\to \infty$ under the null hypothesis, where $\hat{\tau}(t)=g'\{\hat{S}(t)\}\hat{\sigma}(t)$, and $z_p=\Phi^{-1}(p)$ is a standard normal quantile with a probability $p$.
Under the alternative hypothesis $Z$ is approximately distributed as $N(\epsilon\sqrt{n}/\tau_1, 1)$ for large $n$, then the power is approximately given by
\[
\mathrm{P}\left(Z>z_{1-\alpha} \,\middle|\, H_1 \right)
\approx
\Phi\left(-z_{1-\alpha}+\frac{\epsilon\sqrt{n}}{\tau_1} \right),
\]
where $\tau_j^2= g'\{S_j(t)\}^2 \sigma_j^2(t)$ for $j=0,1$ are the asymptotic variances under the null and alternative hypotheses.
To achieve the power of $1-\beta$, we set
\[
1-\beta = \mathrm{\Phi} \left(-z_{1-\alpha}+\frac{\epsilon\sqrt{n_1}}{\tau_1} \right).
\]
The above result lead a sample size formula,
\begin{equation}
	\label{equ:3}
	n_1=\left\{ \frac{\tau_1(z_{1-\alpha}+z_{1-\beta})}{\epsilon} \right\}^2.
\end{equation}
We apply this formula to the transformations, $g\{S(t)\}$, mentioned in Section \ref{sec:2.1}.

\subsection{Existing method to determine sample size}
\label{sec:2.2}

Practically, there is a method to determine sample size \citep{CRAB2002} that has been frequently used in one sample survival studies.
This method is based on the log transformation, $g\{S(t)\}=\log S(t)$.
As noted above, the log transformation has quite poor performance with small-sample sizes \citep{Borgan1990}.
To estimate the sample size, the formula
\begin{equation}
\label{equ:2}
n_2 = \left( \frac{\tau_1 z_{1-\alpha}+\tau_0 z_{1-\beta}}{\epsilon} \right)^2,
\end{equation}
is used in the existing method \citep{CRAB2002}.
By formula (\ref{equ:2}), it follows that
\[
1-\beta=\Phi\left( -\frac{\tau_1}{\tau_0}z_{1-\alpha} + \frac{\epsilon\sqrt{n_2}}{\tau_0} \right).
\]
Therefore, this sample size formula is based on the approximation,
\[
\begin{split}
\mathrm{P}\left(Z>z_{1-\alpha} \,\middle|\, H_1 \right)
&\approx
\Phi\left(-\frac{\tau_1}{\tau_0}z_{1-\alpha} + \frac{\epsilon\sqrt{n}}{\tau_0} \right)
\\&=
\mathrm{P}\left(\frac{\tau_0 W + \epsilon \sqrt{n}}{\tau_1} >z_{1-\alpha} \right),
\end{split}
\]
where $W \sim N(0,1)$.
However, since $\tau_0 \ne \tau_1$, $(\tau_0 W + \epsilon \sqrt{n})/\tau_1$ does not distributed as $N(\epsilon\sqrt{n}/\tau_1, 1)$ but as $N(\epsilon\sqrt{n}/\tau_1, \tau_0^2/\tau_1^2)$ under the alternative hypothesis.
This approximation is inappropriate.
The existing method may be likely designed to compensate for the poor performance of the log transformation.
These problems would possibly influence the accuracy of the results of the sample size calculations.

By the definitions of formulae (2) and (3), $n_1=\{(\tau_1 z_{1-\alpha}+\tau_1 z_{1-\beta})/\epsilon\}^2$ and $n_2=\{(\tau_1 z_{1-\alpha}+\tau_0 z_{1-\beta})/\epsilon\}^2$.
If the type-I error rate is properly controlled, then $n_2$ will be over-estimated when $\tau_0 > \tau_1$ and $n_2$ will be under-estimated when $\tau_0 < \tau_1$.
For illustrative purposes, values of $\tau_0 / \tau_1$ are shown in Table \ref{tab:0} under a simple condition as follows: there is no random censoring and the analysis time $t$ is smaller than the follow-up time $b$.
In particular, values of $\tau_0 / \tau_1$ for the log transformation is greater than 1 (i.e., over-estimation) when $S_0 (t) < S_1 (t)$, and less than 1 (i.e., under-estimation) when $S_0 (t) > S_1 (t)$.
The same applies to power (i.e., over-estimation and under-estimation directly correspond over-power and under-power, respectively), but note that assuming the type-I error rate is well controlled.

\begin{table}[h] 
\centering
\caption{Values of $\tau_0/\tau_1$ under no random censoring and $t < b$.}
\label{tab:0}
\begin{threeparttable}
\begin{tabular}[t]{lcccccc}
\toprule
$S_0(t)$ & $S_1(t)$ & \multicolumn{5}{c}{$\tau_0/\tau_1$} \\
 &  & identity & log & log-log & logit & arcsin \\ \midrule
0.1 & 0.2 & 0.75 & 1.50 & 1.05 & 1.33 & 1.00 \\
0.3 & 0.4 & 0.94 & 1.25 & 0.95 & 1.07 & 1.00 \\
0.5 & 0.6 & 1.02 & 1.22 & 0.90 & 0.98 & 1.00 \\
0.7 & 0.8 & 1.15 & 1.31 & 0.82 & 0.87 & 1.00 \\
0.8 & 0.7 & 0.87 & 0.76 & 1.22 & 1.15 & 1.00 \\
0.6 & 0.5 & 0.98 & 0.82 & 1.11 & 1.02 & 1.00 \\
0.4 & 0.3 & 1.07 & 0.80 & 1.05 & 0.94 & 1.00 \\
0.2 & 0.1 & 1.33 & 0.67 & 0.95 & 0.75 & 1.00 \\ \bottomrule
\end{tabular}
\end{threeparttable}
\end{table}

\subsection{Calculating sample size for continuous and differentiable survival functions}
\label{sec:2.4}
In this sub-section, we consider the numerical computation of survival sample size formulae.
Suppose that $S(t)$ is a parametric continuously differentiable function, with uniform accrual over time, and no loss to follow-up, where $t$ is the analysis time.
Moreover, it is assumed that patients are uniformly recruited during an accrual time $a$, follow-up time is $b$, and the total time is $c=a+b$.
Then, the survival function (i.e., the complementary cumulative distribution function) of the censoring random variable $U$ is given as,
\[
\mathrm{P}(U>t)=
\begin{cases}
1 & 0 < t \leq b\\
a^{-1}(c-t) & b<t\leq c\\
\end{cases},
\]
and the asymptotic variance of the Kaplan--Meier estimator is given by
\[
\sigma^2(t)=
\begin{cases}
S^2(t) \int_0^t \frac{\mathrm{d}\Lambda(s)}{S(s)} & 0 < t \leq b\\
S^2(t) \int_0^b \frac{\mathrm{d}\Lambda(s)}{S(s)}+S^2(t) \int_b^t \frac{a}{(c-s)S(s)}\mathrm{d}\Lambda(s) & b<t\leq c\\
\end{cases},
\]
Since $S(t)$ is continuous and differentiable, if $0 <t \leq b$ (i.e., $\mathrm{P}(U>t)=1$), then
\[
\begin{split}
\int_0^t & \frac{\mathrm{d}\Lambda(s)}{S(s)}
=
\int_0^t -\frac{\mathrm{d}S(s)}{\mathrm{d}s} \frac{1}{\{S(s)\}^2} \mathrm{d}s
=
\left[-\{S(s)\}^{-1} \right]_0^t+
\int_0^t -2\frac{\mathrm{d}S(s)}{\mathrm{d}s} \frac{1}{\{S(s)\}^2} \mathrm{d}s \\
\Leftrightarrow&
\int_0^t -\frac{\mathrm{d}S(s)}{\mathrm{d}s} \frac{1}{\{S(s)\}^2} \mathrm{d}s
=
-\left[-\{S(s)\}^{-1} \right]_0^t
=
-\left\{ -\frac{1}{S(t)}+\frac{1}{S(0)} \right\}=
S^{-1}(t)-1,
\end{split}
\]
by the definition of the survival function, $S(0)=1$.
In this case, we get
\[
\sigma^2(t)=
S^2(t)\{S^{-1}(t)-1\}=
S(t)\{1-S(t)\}.
\]
If the censoring probability, $\mathrm{P}(U>t)$, does not depend on $t$ (i.e., $0<t\leq b$), the asymptotic variance corresponds to the variance of a Binomial distribution with the probability $p=S(t)$.
Thus, the proposed method can be viewed as a natural extension of a Binomial sample size method.
For the arcsine square-root transformation, the asymptotic variance is $\tau^2(t)=g'\{S(t)\}^2 \sigma^2(t)=[4S(t)\{1-S(t)\}]^{-1}S(t)\{1-S(t)\}=1/4$.
It can be considered as a variance stabilizing transformation and a theoretical advantage of the arcsine square-root transformation.

Now consider a more complicated case of $b<t \leq c$.
In this case, the survival function of the censoring random variable $U$ is a function of $t$ (i.e., $\mathrm{P}(U>t)=a^{-1}(c-t)$), and numerical integration is needed to compute the asymptotic variance.
For example, assuming an exponential survival time distribution, $S(t)=\exp(-\lambda t)$, the asymptotic variance can be calculated by
\[
\sigma^2(t)=\exp(-2 \lambda t)
\left[\{\exp(\lambda b)-1\}+
\int_b^t \frac{a \lambda}{(c-s) \exp(-\lambda s)} \mathrm{d}s
\right].
\]
Assuming an exponential survival function $S(t)=\exp(-\lambda t)$, the hazard $\lambda$ is constant over time and can be calculated based on the relation $\lambda=-\log S(t)/t$.
Using the approach of Brookmeyer--Crowley \citep{Brookmeyer1982}, the hazard can also be calculated based on the relation $\lambda=-\log(0.5)/M$, where $M$ is median survival time.
A web application implementing the proposed method that assumes an exponential survival distribution is available at the first author's website ({\tt http://nshi.jp/en/js{\slash}onesurvyr/} and {\tt http://nshi.jp/en/js/onesurvmst/}).

\section{Simulations}
\label{sec:3}
We conducted two sets of simulation studies to evaluate the performances of the existing and proposed methods described in Section 2.
In Simulation 1: type I error rate, we consider whether the transformed Kaplan--Meier estimators can control type I error rate to a nominal level.
In Simulation 2: power, we consider whether the existing and proposed sample size formulae can control power to a prescribed value.

\subsection{Simulation 1: type I error rate of the transformed Kaplan--Meier estimators}
\label{sec:3.1}
First, we re-evaluated the type I error rates of tests based on the Kaplan--Meier estimator with transformations (i.e., the identity, log, log-minus-log, logit, arcsine square-root transformations).
One-sided hypotheses, $H_0: \epsilon \leq 0$, and $H_1: \epsilon > 0$, were considered.

We generated the survival time $X_i$ of the $i$-th observation for $i=1, \ldots, n$ from the exponential and Weibull distributions, $\exp(\lambda)$, $\mathrm{Weibull}(\lambda, k=0.5)$, or $\mathrm{Weibull}(\lambda, k=2)$.
The analysis time was set to $t=12$ months, accrual time was set to $a=24$ months, and follow-up time was set to $b=12$ months.
The null and alternative survival probabilities were set to $S_0(t)=S_1(t)=0.1,0.2,\ldots,0.9$.
For Weibull distributions, the probability density and survival functions are defined as $f(t)=\lambda k(\lambda t)^{k-1} \exp\{-(\lambda t)^k\}$ and $S(t)= \exp\{-(\lambda t)^k\}$, so the hazard function was calculated from $\lambda=\{-\log S(t)\}^{1/k}/t$.
We considered cases of type I censoring both with and without random censoring.
For type I censoring, we generated the observed accrual time of the $i$-th observation $a_i$ from a uniform distribution $U(0,a)$ and if $a_i+X_i>a+b$ then $i$-th observation was treated as a censored observation.
For random censoring, we generated the censoring time of the $i$-th observation $U_i$ from exponential and Weibull distributions, $\exp(4 \lambda)$, $\mathrm{Weibull}(4^{1/k} \lambda, k=0.5)$, or $\mathrm{Weibull}(4^{1/k} \lambda, k=2)$ (i.e., from the same distribution with different parameters as the survival time distribution), and if $U_i>X_i$ then the $i$-th observation was treated as a censored observation.
The hazard functions with $4\lambda$ or $4^{1/k}\lambda$ correspond to an approximately $20\%$ censoring rate.
The one-sided type I error rate was set to $\alpha=5\%$, and the sample size was set to $n=25,50$, or $100$.
For each setting, we simulated 1,000,000 replications.
To assess the performance, the empirical type I error was estimated by
\begin{equation}
\label{equ:phat}
\hat{P}=\frac{\{\# \text{ of replications with confidence intervals not covering the null hypothesis } S_0(t)\}}{\{\# \text{ of replications } (1,000,000)\}}.
\end{equation}

\begin{table}[p] 
\centering
\small
\caption{Simulation results for type I error rate; no random censoring was assumed.}
\label{tab:1}
\begin{threeparttable}
\begin{tabular}{crrrrrrr}
\toprule
\multicolumn{1}{c}{Distribution} & \multicolumn{1}{c}{Sample size} & \multicolumn{1}{c}{$S_0(t)$} & \multicolumn{1}{c}{identity} & \multicolumn{1}{c}{log} & \multicolumn{1}{c}{log-log} & \multicolumn{1}{c}{logit} & \multicolumn{1}{c}{arcsine} \\ \midrule
Exponential & 25 & 0.10 & 0.010 & \textbf{0.098} & 0.033 & 0.033 & 0.033 \\
 &  & 0.30 & 0.044 & \textbf{0.098} & 0.044 & 0.044 & 0.044 \\
 &  & 0.50 & 0.054 & \textbf{0.115} & 0.054 & 0.054 & 0.054 \\
 &  & 0.70 & \textbf{0.091} & \textbf{0.091} & 0.033 & 0.033 & \textbf{0.091} \\
 &  & 0.90 & \textbf{0.072} & \textbf{0.072} & \textbf{0.072} & \textbf{0.072} & \textbf{0.072} \\
 & 50 & 0.10 & 0.024 & 0.058 & 0.058 & 0.058 & 0.058 \\
 &  & 0.30 & 0.047 & \textbf{0.084} & 0.047 & 0.047 & 0.047 \\
 &  & 0.50 & 0.059 & 0.059 & 0.032 & 0.059 & 0.059 \\
 &  & 0.70 & \textbf{0.078} & \textbf{0.078} & 0.040 & 0.040 & 0.040 \\
 &  & 0.90 & \textbf{0.112} & \textbf{0.112} & 0.033 & 0.033 & \textbf{0.112} \\
 & 100 & 0.10 & 0.021 & \textbf{0.072} & 0.040 & \textbf{0.072} & 0.040 \\
 &  & 0.30 & 0.053 & 0.053 & 0.053 & 0.053 & 0.053 \\
 &  & 0.50 & 0.044 & 0.066 & 0.044 & 0.044 & 0.044 \\
 &  & 0.70 & \textbf{0.075} & \textbf{0.075} & 0.048 & 0.048 & 0.048 \\
 &  & 0.90 & \textbf{0.117} & \textbf{0.117} & 0.024 & 0.024 & 0.057 \\ \midrule
$\mathrm{Weibull}(\lambda, k=0.5)$ & 25 & 0.10 & 0.009 & \textbf{0.098} & 0.033 & 0.033 & 0.033 \\
 &  & 0.30 & 0.044 & \textbf{0.098} & 0.044 & 0.044 & 0.044 \\
 &  & 0.50 & 0.054 & \textbf{0.115} & 0.054 & 0.054 & 0.054 \\
 &  & 0.70 & \textbf{0.090} & \textbf{0.090} & 0.033 & 0.033 & \textbf{0.090} \\
 &  & 0.90 & \textbf{0.072} & \textbf{0.072} & \textbf{0.072} & \textbf{0.072} & \textbf{0.072} \\ 
 & 50 & 0.10 & 0.024 & 0.058 & 0.058 & 0.058 & 0.058 \\
 &  & 0.30 & 0.048 & \textbf{0.085} & 0.048 & 0.048 & 0.048 \\
 &  & 0.50 & 0.060 & 0.060 & 0.032 & 0.060 & 0.060 \\
 &  & 0.70 & \textbf{0.078} & \textbf{0.078} & 0.040 & 0.040 & 0.040 \\
 &  & 0.90 & \textbf{0.112} & \textbf{0.112} & 0.034 & 0.034 & \textbf{0.112} \\
 & 100 & 0.10 & 0.020 & \textbf{0.072} & 0.040 & \textbf{0.072} & 0.040 \\
 &  & 0.30 & 0.053 & 0.053 & 0.053 & 0.053 & 0.053 \\
 &  & 0.50 & 0.044 & 0.066 & 0.044 & 0.044 & 0.044 \\
 &  & 0.70 & \textbf{0.075} & \textbf{0.075} & 0.048 & 0.048 & 0.048 \\
 &  & 0.90 & \textbf{0.117} & \textbf{0.117} & 0.024 & 0.024 & 0.058 \\ \midrule
$\mathrm{Weibull}(\lambda, k=2)$ & 25 & 0.10 & 0.010 & \textbf{0.098} & 0.033 & 0.033 & 0.033 \\
 &  & 0.30 & 0.044 & \textbf{0.098} & 0.044 & 0.044 & 0.044 \\
 &  & 0.50 & 0.054 & \textbf{0.115} & 0.054 & 0.054 & 0.054 \\
 &  & 0.70 & \textbf{0.091} & \textbf{0.091} & 0.033 & 0.033 & \textbf{0.091} \\
 &  & 0.90 & \textbf{0.072} & \textbf{0.072} & \textbf{0.072} & \textbf{0.072} & \textbf{0.072} \\
 & 50 & 0.10 & 0.024 & 0.058 & 0.058 & 0.058 & 0.058 \\
 &  & 0.30 & 0.048 & \textbf{0.085} & 0.048 & 0.048 & 0.048 \\
 &  & 0.50 & 0.060 & 0.060 & 0.033 & 0.060 & 0.060 \\
 &  & 0.70 & \textbf{0.079} & \textbf{0.079} & 0.040 & 0.040 & 0.040 \\
 &  & 0.90 & \textbf{0.112} & \textbf{0.112} & 0.034 & 0.034 & \textbf{0.112} \\
 & 100 & 0.10 & 0.020 & \textbf{0.072} & 0.040 & \textbf{0.072} & 0.040 \\
 &  & 0.30 & 0.053 & 0.053 & 0.053 & 0.053 & 0.053 \\
 &  & 0.50 & 0.044 & 0.067 & 0.044 & 0.044 & 0.044 \\
 &  & 0.70 & \textbf{0.076} & \textbf{0.076} & 0.048 & 0.048 & 0.048 \\
 &  & 0.90 & \textbf{0.118} & \textbf{0.118} & 0.024 & 0.024 & 0.058 \\ \bottomrule
\end{tabular}

\begin{tablenotes}
\item Values not within a range of type I error rate less than $\alpha=0.05$ plus $0.02$ are highlighted in \textbf{bold}.
\end{tablenotes}
\end{threeparttable}
\end{table}

The type I error rates for the case with no random censoring are shown in Table \ref{tab:1}.
The type I error rate for the log-minus-log was closest to the nominal level of $5\%$, except for a few of the cases where $S_0(t)=0.9$.
The logit and arcsine square-root transformations were generally quite close to the nominal level of $5\%$, except for a few cases when $S_0(t)=0.1,0.7$, or $0.9$.
The type I error rates for the identity transformation were smaller than the nominal value of $5\%$ when $S_0(t)=0.1$, and were larger than the nominal value when $S_0(t)=0.7$ and $0.9$.
In the worst cases, the error rates were about $1\%$, which is highly conservative, or about $10\%$, which is highly inflated.
The type I error rates for the log transformation were larger than the nominal value in all cases.
They were inflated in almost all cases, and the error rate was greater than $10\%$ in the worst case.

Due to the non-parametric nature of the Kaplan--Meier estimator, exponential and Weibull survival time distributions give similar results.
Moreover, type I censoring, with or without random censoring, also gives similar results (see Supporting Information Table S1), since the Kaplan--Meier estimator is valid under both models.

In summary, the Kaplan--Meier estimator with log-minus-log, logit, arcsine square-root transformations controlled type-I error rates, and the type-I error rates of the Kaplan--Meier estimator with identity and log transformations were conservative or inflated in many cases.

\subsection{Simulation 2: power of the sample size methods}
\label{sec:3.2}

In this section, we evaluated the power of the existing method for determining sample size described in Section \ref{sec:2.2} and the proposed methods described in Section \ref{sec:2.3}.

We generated the survival time $X_i$ of the $i$-th observation for $i=1, \ldots, n$ from the exponential and Weibull distributions, $\exp(\lambda)$, $\mathrm{Weibull}(\lambda, k=0.5)$, or $\mathrm{Weibull}(\lambda, k=2)$.
The analysis time was set to $t=12$ months, accrual time was set to $a=24$ months, and follow-up time was set to $b=6$ or $12$ months.
The null and alternative survival probabilities were set to $S_0(t)=0.1,0.4$, or $0.7$ and $S_1(t)-S_0(t)=0.1$.
We also considered both cases of type I censoring with or without random censoring.
The one-sided type I error rate was set to $\alpha=5\%$, or $10\%$, and the power was set to $1-\beta=80\%$, or $90\%$.
The sample sizes were calculated from equation (\ref{equ:2}) (i.e., the existing method) and (\ref{equ:3}) for the log transformation, and only from equation (\ref{equ:3}) for the other transformations.
The same transformation was used to determine sample sizes and estimate intervals for the survival function.
For each setting, we simulated 1,000,000 replications.
To assess the performance of the sample size methods, the empirical power under the calculated sample size was estimated by (\ref{equ:phat}).

The power for the $\alpha=0.05$, $1-\beta=0.8$, and exponential survival distribution case is shown in Table \ref{tab:2}.
The sample sizes based on each method were substantially different.
The largest difference among transformations was between the identity and log transformations with $b=6$, $S_0(t)=0.1$, and with random censoring (difference of 69).
The power for the proposed method with the arcsine transformation was closest to the prescribed  values of $80\%$ or $90\%$.
The existing method (i.e., formula (\ref{equ:2}) with the log transformation) had a good performance, but this result is not consistent with the results of the type I error rate.
The power for the proposed method with the identity transformation was larger than the prescribed values when $S_0(t)=0.1$, and tended to be smaller than the prescribed values when $S_0(t)=0.7$.
The power for the proposed method with the log transformation was substantially smaller than the prescribed values in all cases.
This result reflects the type I error rate inflation of the log transformation.
The power values for the proposed method with the log-minus-log and logit transformation were smaller than the prescribed values when $S_0(t)=0.1$ and were larger than the prescribed values when $S_0(t)=0.7$.

Similar to the case of the type I error rates, the exponential and Weibull survival time distributions gave similar results, and type I censoring with or without random censoring also gave similar results (see Supporting Information Tables S2--S7).

In summary, the proposed method with the arcsine square-root transformation was able to maintain the prescribed power in almost all cases.
The existing method was second best, but since the type I error rate of the log transformation was inflated, it is not acceptable for sample size calculation.
The other methods were not able to control power to the prescribed levels.
Moreover, it is dangerous to use the identity and log transformations, since the type I errors were inflated.

Source code to reproduce the results is available as Supporting Information.

\begin{table}[p] 
\centering
\footnotesize
\caption{Simulation results for power simulation ($\alpha=0.05$ and $1-\beta=0.8$); exponential survival distribution was assumed.}
\label{tab:2}
\hspace*{-1.5cm}\begin{threeparttable}
\begin{tabular}{crrrrrrrrrrrrrrr}
\toprule
\multicolumn{1}{c}{Random} & \multicolumn{1}{c}{$b$} & \multicolumn{1}{c}{$S_0(t)$} & \multicolumn{1}{c}{$S_1(t)$} & \multicolumn{6}{c}{Sample size} & \multicolumn{6}{c}{Empirical power} \\
\multicolumn{1}{c}{censoring} &  &  &  & \multicolumn{1}{c}{ident.} & \multicolumn{1}{c}{log} & \multicolumn{1}{c}{log (3)} & \multicolumn{1}{c}{log-log} & \multicolumn{1}{c}{logit} & \multicolumn{1}{c}{arcsin} & \multicolumn{1}{c}{ident.} & \multicolumn{1}{c}{log} & \multicolumn{1}{c}{log (3)} & \multicolumn{1}{c}{log-log} & \multicolumn{1}{c}{logit} & \multicolumn{1}{c}{arcsin} \\ \midrule
\multicolumn{1}{c}{without} & 12 & 0.1 & 0.2 & 99 & 52 & 71 & 75 & 59 & 77 & \textbf{0.861} & \textbf{0.739} & 0.786 & \textbf{0.761} & \textbf{0.769} & 0.794 \\
 &  & 0.4 & 0.5 & 155 & 125 & 144 & 166 & 151 & 153 & 0.789 & \textbf{0.762} & 0.820 & 0.803 & 0.792 & 0.791 \\
 &  & 0.7 & 0.8 & 99 & 87 & 106 & 142 & 134 & 115 & \textbf{0.755} & \textbf{0.719} & 0.791 & \textbf{0.857} & \textbf{0.845} & 0.795 \\
 & 6 & 0.1 & 0.2 & 111 & 58 & 80 & 84 & 66 & 86 & \textbf{0.832} & \textbf{0.716} & 0.814 & 0.784 & \textbf{0.739} & 0.785 \\
 &  & 0.4 & 0.5 & 170 & 136 & 158 & 181 & 165 & 167 & 0.801 & \textbf{0.760} & 0.808 & 0.815 & 0.798 & 0.799 \\
 &  & 0.7 & 0.8 & 107 & 94 & 115 & 153 & 144 & 125 & 0.777 & \textbf{0.755} & 0.818 & \textbf{0.850} & \textbf{0.838} & 0.809 \\
\multicolumn{1}{c}{with} & 12 & 0.1 & 0.2 & 129 & 67 & 98 & 97 & 77 & 100 & \textbf{0.832} & \textbf{0.713} & 0.829 & 0.782 & \textbf{0.740} & 0.785 \\
 &  & 0.4 & 0.5 & 171 & 137 & 161 & 183 & 166 & 169 & 0.802 & \textbf{0.761} & 0.813 & 0.818 & 0.798 & 0.801 \\
 &  & 0.7 & 0.8 & 102 & 90 & 110 & 146 & 137 & 119 & 0.779 & \textbf{0.762} & 0.822 & \textbf{0.851} & \textbf{0.839} & 0.811 \\
 & 6 & 0.1 & 0.2 & 145 & 76 & 111 & 109 & 87 & 113 & \textbf{0.861} & \textbf{0.747} & \textbf{0.856} & 0.812 & 0.773 & 0.817 \\
 &  & 0.4 & 0.5 & 188 & 151 & 178 & 201 & 183 & 185 & \textbf{0.841} & 0.801 & \textbf{0.852} & \textbf{0.855} & \textbf{0.839} & \textbf{0.839} \\
 &  & 0.7 & 0.8 & 111 & 97 & 119 & 158 & 149 & 129 & 0.804 & 0.781 & \textbf{0.843} & \textbf{0.874} & \textbf{0.864} & \textbf{0.835} \\ \bottomrule
\end{tabular}

\begin{tablenotes}
\item Values not within a range of $1-\beta=0.8$, plus or minus $0.03$ are highlighted in \textbf{bold}.
\item ident.: the identity transformation based on equation (\ref{equ:3}); log: the log transformation based on equation (\ref{equ:3}); log (\ref{equ:2}): the log transformation based on equation (\ref{equ:2}), i.e., the existing method; log-log: the log-log transformation based on equation (\ref{equ:3}); logit: the logit transformation based on equation (\ref{equ:3}); arcsin: the arcsine square-root transformation based on equation (\ref{equ:3}).
\end{tablenotes}
\end{threeparttable}
\end{table}

\begin{table}[p] 
\centering
\scriptsize
\caption{Power simulation results based on three clinical studies; exponential survival distribution and no random censoring were assumed.}
\label{tab:3}
\hspace*{-0.5cm}\begin{threeparttable}
\begin{tabular}{lrrrrrrrrrrrrrr}
\toprule
\multicolumn{1}{c}{Study} & \multicolumn{1}{c}{$1-\beta$} & \multicolumn{7}{c}{Sample size} & \multicolumn{6}{c}{Empirical power} \\
 &  & \multicolumn{1}{c}{actual} & \multicolumn{1}{c}{ident.} & \multicolumn{1}{c}{log} & \multicolumn{1}{c}{log (3)} & \multicolumn{1}{c}{log-log} & \multicolumn{1}{c}{logit} & \multicolumn{1}{c}{arcsin} & \multicolumn{1}{c}{ident.} & \multicolumn{1}{c}{log} & \multicolumn{1}{c}{log (3)} & \multicolumn{1}{c}{log-log} & \multicolumn{1}{c}{logit} & \multicolumn{1}{c}{arcsin} \\ \midrule
(i) & 0.90 &
  50 & 45 & 33 & 50 & 66 & 57 & 51 &
  0.901 & \textbf{0.839} & 0.915 & \textbf{0.935} & \textbf{0.938} & 0.899 \\
(ii) & 0.82 &
  70 & 73 & 53 & 68 & 83 & 73 & 73 &
  0.805 & \textbf{0.768} & 0.830 & \textbf{0.872} & 0.805 & 0.805 \\
(iii) & 0.90 &
  37 & 35 & 18 & 32 & 38 & 29 & 32 &
  0.913 & \textbf{0.761} & \textbf{0.945} & \textbf{0.929} & \textbf{0.868} & 0.893 \\ \bottomrule
\end{tabular}

\begin{tablenotes}
\item Values not within a range of $1-\beta=0.90, 0.82$, or $0.90$, plus or minus 0.02 are highlighted in \textbf{bold}.
\item actual: planned sample sizes in the actual studies; ident.: the identity transformation based on equation (\ref{equ:3}); log: the log transformation based on equation (\ref{equ:3}); log (\ref{equ:2}): the log transformation based on equation (\ref{equ:2}), i.e., the existing method; log-log: the log-log transformation based on equation (\ref{equ:3}); logit: the logit transformation based on equation (\ref{equ:3}); arcsin: the arcsine square-root transformation based on equation (\ref{equ:3}).
\end{tablenotes}
\end{threeparttable}
\end{table}

\section{Applications to clinical trials}
\label{sec:4}

We applied the sample size methods to data from three clinical trials \citep{Spigel2013,Lai2011,Grignani2015}.
In reference to the three trials, the simulation conditions for each study were determined to evaluate the empirical performance of the methods for determining sample size under realistic scenarios:
\begin{description}
\item[(i) \citep{Spigel2013}: ]
The primary endpoint was progression free survival at the analysis time $t=3$ months, accrual time was set to $a=22$ months, and follow-up time was set to $b=4$ months.
The null and alternative survival probabilities were set to $S_0(t)=0.50$ and $S_1(t)=0.70$.
The one-sided type I error rate was set to $\alpha=5\%$, and the power was set to $1-\beta=90\%$.
In the actual study, 50 assessable participants were planned using the existing method \cite{CRAB2002}.
\item[(ii) \citep{Lai2011}: ]
The primary endpoint was overall survival at the analysis time $t=18$ months, accrual time was set to $a=27$ months, and follow-up time was set to $b=18$ months.
The null and alternative survival probabilities were set to $S_0(t)=0.40$ and $S_1(t)=0.55$.
The one-sided type I error rate was set to $\alpha=5\%$, and the power was set to $1-\beta=82\%$.
In the actual study, 70 patients were enrolled, and the method used to determine sample size was not reported in the article.
\item[(iii) \citep{Grignani2015}: ]
The primary endpoint was progression free survival at the analysis time $t=6$ months, accrual time was set to $a=23$ months, and follow-up time was set to $b=6$ months.
The null and alternative survival probabilities were set to $S_0(t)=0.25$ and $S_1(t)=0.50$.
The one-sided type I error rate was set to $\alpha=5\%$, and the power was set to $1-\beta=90\%$.
In the actual study, 37 assessable participants were planned using a Simon 2-stage optimal design.
\end{description}

We generated the survival time $X_i$ of the $i$-th observation for $i=1, \ldots, n$ from an exponential distribution, $\exp(\lambda)$.
We considered type I censoring without random censoring.
Other settings were the same as those described in Section \ref{sec:3.2}.

The power values for the exponential survival distribution and no random censoring case are shown in Table \ref{tab:3}.
The results described in Tables \ref{tab:2} and \ref{tab:3} are similar.
The power for the proposed method with the arcsine transformation was closest to the prescribed values.
The power for the proposed method with the identity transformation and the existing method, which is slightly inflated in study (iii), were next best.
The power for the proposed method with the log transformation was substantially smaller than the prescribed values in all cases.
The power values for the proposed method with the log-minus-log and logit transformations were smaller than the prescribed values in some cases and larger in others.
There are some differences between the sample size of the actual studies and those described in Table \ref{tab:3}.
It is notable that the study (iii) applied a Binomial sample size method with an interim analysis.
We shall discuss this in Section \ref{sec:5}.

\section{Discussions}
\label{sec:5}

Since the asymptotic distribution of the Kaplan--Meier estimator is a well-known result, methods for determining sample size have generally not been discussed and have remained unclear.
The existing method for determining sample size has a theoretical problem because it uses an inappropriate standard normal approximation.
We have presented an alternative sample size formula based on an appropriate approximation.

Our simulation results confirmed previous results \citep{Borgan1990} that the type I error rates of the identity and log transformation are inflated or conservative, and we found that the power of the proposed method using the arcsine square-root transformation was closest to the prescribed power.
Several survival time distributions and censoring types gave similar results.
The existing method based on formula (\ref{equ:2}), which uses the log transformation, seemed second best, but the result was not consistent with its type I error rate.
Since the type I error rate of the log transformation was highly inflated, it is expected that  calculated sample sizes are small.
As a result, the power of the existing method should be smaller than the prescribed values.
However, the calculated sample sizes of the formula (\ref{equ:2}) are slightly larger than expected, and it is a cause of the inconsistency.
The log transformation leads to serious type I error inflation, so this method is not recommended for most applications.

The single-arm survival design is often used in early-phase oncology studies and clinical trials for relatively severe or rare diseases such as metastatic colorectal cancer, advanced hepatocellular carcinoma, and childhood disease.
False positive findings due to the type I error inflation of the existing method could lead to an unnecessary large-scale phase III trial, and it could raise serious ethical issues.
By contrast, the proposed method using the arcsine square-root transformation achieved a prescribed level of power, and the type I error rate of the arcsine square-root transformation was not inflated in most cases.
Alternatively, the proposed method using the log-minus-log transformation would be an option to strictly control the type I error rate.
Therefore, considering such ethical issues, we recommend this proposed method in practice because of its accuracy.

We have described related sample size formula for the cases of continuous and differentiable survival functions.
If the censoring probability does not depend on $t$, then the asymptotic variance corresponds to the variance of a Binomial distribution with the probability $p=S(t)$.
The proposed method can be viewed as a natural extension of Binomial sample size methods.
Sample size methods based on a Binomial distribution such as the Fleming single-stage design \citep{Fleming1982} and Simon’s two stage design \citep{Simon1989} are often used in oncology studies.
Using the assumptions of censoring probability described above, the use of the arcsine square-root transformation is also recommended for Fleming single-stage designs.
Although the Simon's two stage design considers an interim analysis at a time $t_m<t$, the assumption does not hold at $t_m$ in general.
Particular attention is required when applying Binomial sample size methods in single-arm survival studies.
Moreover, we showed that the arcsine square-root transformation can be considered as a variance stabilizing transformation when the censoring probability does not depend on $t$.
The simulation results were consistent with the theoretical advantages of the arcsine square-root transformation.
Thomas and Grunkemeier \cite{Thomas1975} also suggested the use of the arcsine-square root transformation to improve the small sample properties of confidence intervals for survival functions.
The confidence interval based on the arcsine square-root transformation has an acceptable performance even with small sample sizes \citep{Borgan1990,Aalen2008}.

We mainly discussed about sample size formula for the survival proportion at time $t$ as a summary measure.
Other measures such as the median survival time and mean survival time are also commonly used.
As noted in Section \ref{sec:2.4}, the proposed method can be easily extended to median survival time by using the Brookmeyer--Crowley \cite{Brookmeyer1982} approach.
Further research will be needed to extend to restricted mean survival time \cite{Irwin1949,Royston2011}.
It will also be necessary to carefully consider the issue of small samples \cite{Lawrence2019,Horiguchi2020}. 

The default settings used in statistical software should be considered.
{\tt S-PLUS} and {\tt R} use log transformation as default, and {\tt SAS} and {\tt Stata} use the log-minus-log transformation \citep{Barker2009}.
Therefore, researchers should be careful when using the confidence intervals for the Kaplan--Meier estimator, especially when using {\tt S-PLUS} and {\tt R}, because the type I error rate will be inflated.

In conclusion, we showed that the proposed sample size calculation method using the arcsine square-root  and log-minus-log transformations work well for single-arm survival studies, even with small sample sizes which are typical in single-arm trials, a case of a non-exponential survival function, and a case of a complex censoring distribution.

\subsection*{Acknowledgements}
The authors would like to thank the associate editor and three reviewers for very insightful and constructive comments that substantially improved the article.
The first author deeply thanks Dr. Junichi Asano at Pharmaceuticals and Medical Devices Agency and Dr. Hiroyuki Sato at Tokyo Medical and Dental University for their valuable advice and suggestions.
This work was supported by JSPS KAKENHI Grant Number 16K16014.


\clearpage
\appendix

\section*{Supplementary Tables}
\setcounter{table}{0}
\def\thetable{S\arabic{table}}
\begin{enumerate}[align=left]
\item[\textbf{Table S1}.] Simulation results for type I error rate: random censoring was assumed.
\item[\textbf{Table S2}.] Simulation results for power: exponential survival distribution and no random censoring were assumed.
\item[\textbf{Table S3}.] Simulation results for power: $\mathrm{Weibull}(\lambda, k=0.5)$ survival distribution and no random censoring were assumed.
\item[\textbf{Table S4}.] Simulation results for power: $\mathrm{Weibull}(\lambda, k=2)$ survival distribution and no random censoring were assumed.
\item[\textbf{Table S5}.] Simulation results for power: exponential survival distribution and no random censoring were assumed.
\item[\textbf{Table S6}.] Simulation results for power: $\mathrm{Weibull}(\lambda, k=0.5)$ survival distribution and random censoring were assumed.
\item[\textbf{Table S7}.] Simulation results for power: $\mathrm{Weibull}(\lambda, k=2)$ survival distribution and random censoring were assumed.
\end{enumerate}

\clearpage
\begin{table}[p] 
\centering
\fontsize{13}{13}\selectfont
\caption{Simulation results for type I error rate: random censoring was assumed.}
\label{tab:s1}
\begin{tabular}{crrrrrrr} \hline
Distribution & \multicolumn{1}{c}{Sample size} & \multicolumn{1}{c}{$S_0(t)$} & \multicolumn{1}{c}{identity} & \multicolumn{1}{c}{log} & \multicolumn{1}{c}{log-log} & \multicolumn{1}{c}{logit} & \multicolumn{1}{c}{arcsine} \\ \hline
Exponential & 25 & 0.10 & 0.024 & \textbf{0.084} & 0.046 & 0.070 & 0.045 \\
 &  & 0.30 & 0.044 & \textbf{0.079} & 0.039 & 0.055 & 0.049 \\
 &  & 0.50 & 0.062 & \textbf{0.085} & 0.037 & 0.046 & 0.055 \\
 &  & 0.70 & \textbf{0.096} & \textbf{0.107} & 0.035 & 0.041 & 0.056 \\
 &  & 0.90 & \textbf{0.075} & \textbf{0.075} & \textbf{0.075} & \textbf{0.075} & \textbf{0.075} \\
 & 50 & 0.10 & 0.026 & \textbf{0.074} & 0.046 & 0.065 & 0.044 \\
 &  & 0.30 & 0.043 & 0.070 & 0.042 & 0.054 & 0.048 \\
 &  & 0.50 & 0.056 & \textbf{0.074} & 0.041 & 0.049 & 0.052 \\
 &  & 0.70 & 0.070 & \textbf{0.086} & 0.039 & 0.046 & 0.056 \\
 &  & 0.90 & \textbf{0.118} & \textbf{0.118} & 0.022 & 0.031 & \textbf{0.104} \\
 & 100 & 0.10 & 0.029 & 0.067 & 0.047 & 0.061 & 0.045 \\
 &  & 0.30 & 0.044 & 0.064 & 0.045 & 0.053 & 0.048 \\
 &  & 0.50 & 0.053 & 0.066 & 0.044 & 0.049 & 0.051 \\
 &  & 0.70 & 0.064 & \textbf{0.072} & 0.042 & 0.046 & 0.055 \\
 &  & 0.90 & \textbf{0.088} & \textbf{0.108} & 0.026 & 0.026 & 0.062 \\ \hline
$\mathrm{Weibull}(\lambda, k=0.5)$ & 25 & 0.10 & 0.024 & \textbf{0.084} & 0.046 & 0.070 & 0.045 \\
 &  & 0.30 & 0.044 & \textbf{0.079} & 0.039 & 0.055 & 0.049 \\
 &  & 0.50 & 0.062 & \textbf{0.085} & 0.037 & 0.046 & 0.055 \\
 &  & 0.70 & \textbf{0.096} & \textbf{0.106} & 0.034 & 0.040 & 0.056 \\
 &  & 0.90 & \textbf{0.075} & \textbf{0.075} & \textbf{0.075} & \textbf{0.075} & \textbf{0.075} \\
 & 50 & 0.10 & 0.025 & \textbf{0.074} & 0.046 & 0.064 & 0.044 \\
 &  & 0.30 & 0.043 & 0.070 & 0.042 & 0.054 & 0.048 \\
 &  & 0.50 & 0.056 & \textbf{0.073} & 0.041 & 0.049 & 0.052 \\
 &  & 0.70 & 0.070 & \textbf{0.085} & 0.039 & 0.046 & 0.056 \\
 &  & 0.90 & \textbf{0.117} & \textbf{0.117} & 0.022 & 0.031 & \textbf{0.104} \\
 & 100 & 0.10 & 0.029 & 0.067 & 0.047 & 0.061 & 0.044 \\
 &  & 0.30 & 0.044 & 0.064 & 0.045 & 0.053 & 0.049 \\
 &  & 0.50 & 0.053 & 0.067 & 0.044 & 0.050 & 0.052 \\
 &  & 0.70 & 0.064 & \textbf{0.072} & 0.042 & 0.046 & 0.055 \\
 &  & 0.90 & \textbf{0.088} & \textbf{0.108} & 0.026 & 0.026 & 0.063 \\ \hline
$\mathrm{Weibull}(\lambda, k=2)$ & 25 & 0.10 & 0.024 & \textbf{0.085} & 0.046 & \textbf{0.071} & 0.045 \\
 &  & 0.30 & 0.044 & \textbf{0.079} & 0.039 & 0.055 & 0.049 \\
 &  & 0.50 & 0.062 & \textbf{0.086} & 0.037 & 0.047 & 0.056 \\
 &  & 0.70 & \textbf{0.096} & \textbf{0.107} & 0.035 & 0.041 & 0.057 \\
 &  & 0.90 & \textbf{0.075} & \textbf{0.075} & \textbf{0.075} & \textbf{0.075} & \textbf{0.075} \\
 & 50 & 0.10 & 0.026 & \textbf{0.073} & 0.046 & 0.064 & 0.044 \\
 &  & 0.30 & 0.043 & \textbf{0.070} & 0.042 & 0.054 & 0.048 \\
 &  & 0.50 & 0.056 & \textbf{0.074} & 0.041 & 0.049 & 0.053 \\
 &  & 0.70 & 0.070 & \textbf{0.086} & 0.039 & 0.046 & 0.056 \\
 &  & 0.90 & \textbf{0.117} & \textbf{0.117} & 0.022 & 0.031 & \textbf{0.103} \\
 & 100 & 0.10 & 0.029 & 0.067 & 0.047 & 0.060 & 0.044 \\
 &  & 0.30 & 0.044 & 0.064 & 0.044 & 0.053 & 0.048 \\
 &  & 0.50 & 0.053 & 0.066 & 0.044 & 0.050 & 0.051 \\
 &  & 0.70 & 0.064 & \textbf{0.072} & 0.043 & 0.047 & 0.055 \\
 &  & 0.90 & \textbf{0.088} & \textbf{0.108} & 0.026 & 0.026 & 0.062 \\ \hline
\multicolumn{8}{l}{\footnotesize Values not within a nominal range of type I error rate less than $\alpha=0.05+0.02$ are highlighted in}\\
\multicolumn{8}{l}{\footnotesize \textbf{bold}.}
\end{tabular}
\end{table}

\clearpage

\begin{landscape}

\begin{table}[p] 
\centering
\fontsize{13}{13}\selectfont
\caption{Simulation results for the power: exponential survival distribution and no random censoring were assumed.}
\label{tab:s2}
\begin{tabular}{lllrrrrrrrrrrrrrr} \hline
\multicolumn{1}{c}{\multirow{2}{*}{$b$}} & \multicolumn{1}{c}{\multirow{2}{*}{$S_0(t)$}} & \multicolumn{1}{c}{\multirow{2}{*}{$S_1(t)$}} & \multicolumn{1}{c}{\multirow{2}{*}{$\alpha$}} & \multicolumn{1}{c}{\multirow{2}{*}{$1-\beta$}} & \multicolumn{6}{c}{Sample size} & \multicolumn{6}{c}{Empirical power} \\
 &  &  &  &  & \multicolumn{1}{c}{ident.} & \multicolumn{1}{c}{log} & \multicolumn{1}{c}{log (2)} & \multicolumn{1}{c}{log-log} & \multicolumn{1}{c}{logit} & \multicolumn{1}{c}{arcsin} & \multicolumn{1}{c}{ident.} & \multicolumn{1}{c}{log} & \multicolumn{1}{c}{log (2)} & \multicolumn{1}{c}{log-log} & \multicolumn{1}{c}{logit} & \multicolumn{1}{c}{arcsin} \\ \hline
\multicolumn{1}{r}{12} & \multicolumn{1}{r}{0.1} & \multicolumn{1}{r}{0.2} & 0.05 & 0.80 & 99 & 52 & 71 & 75 & 59 & 77 & \textbf{0.861} & \textbf{0.739} & 0.786 & \textbf{0.761} & \textbf{0.769} & 0.794 \\
 &  &  & 0.10 & 0.80 & 73 & 38 & 54 & 55 & 43 & 56 & 0.817 & \textbf{0.660} & 0.780 & 0.798 & 0.784 & 0.815 \\
 &  &  & 0.05 & 0.90 & 138 & 72 & 106 & 104 & 82 & 107 & \textbf{0.939} & \textbf{0.802} & 0.921 & 0.907 & \textbf{0.861} & 0.884 \\
 &  &  & 0.10 & 0.90 & 106 & 55 & 86 & 80 & 63 & 82 & 0.921 & \textbf{0.798} & 0.901 & 0.900 & \textbf{0.836} & 0.916 \\ \hline
\multicolumn{1}{r}{12} & \multicolumn{1}{r}{0.4} & \multicolumn{1}{r}{0.5} & 0.05 & 0.80 & 155 & 125 & 144 & 166 & 151 & 153 & 0.789 & \textbf{0.762} & 0.820 & 0.803 & 0.792 & 0.791 \\
 &  &  & 0.10 & 0.80 & 113 & 91 & 108 & 121 & 110 & 112 & 0.826 & \textbf{0.735} & 0.806 & 0.818 & 0.804 & 0.802 \\
 &  &  & 0.05 & 0.90 & 215 & 172 & 208 & 229 & 209 & 212 & 0.890 & 0.874 & 0.906 & 0.907 & 0.893 & 0.904 \\
 &  &  & 0.10 & 0.90 & 165 & 132 & 164 & 176 & 160 & 163 & 0.894 & 0.871 & 0.908 & 0.924 & 0.911 & 0.895 \\ \hline
\multicolumn{1}{r}{12} & \multicolumn{1}{r}{0.7} & \multicolumn{1}{r}{0.8} & 0.05 & 0.80 & 99 & 87 & 106 & 142 & 134 & 115 & \textbf{0.755} & \textbf{0.719} & 0.791 & \textbf{0.857} & \textbf{0.845} & 0.795 \\
 &  &  & 0.10 & 0.80 & 73 & 64 & 80 & 103 & 97 & 84 & 0.805 & 0.803 & \textbf{0.837} & \textbf{0.832} & \textbf{0.851} & 0.773 \\
 &  &  & 0.05 & 0.90 & 138 & 121 & 155 & 196 & 185 & 160 & 0.893 & \textbf{0.836} & \textbf{0.931} & \textbf{0.949} & \textbf{0.938} & 0.898 \\
 &  &  & 0.10 & 0.90 & 106 & 93 & 123 & 150 & 142 & 123 & 0.899 & \textbf{0.845} & 0.906 & \textbf{0.934} & 0.929 & 0.906 \\ \hline 
\multicolumn{1}{r}{6} & \multicolumn{1}{r}{0.1} & \multicolumn{1}{r}{0.2} & 0.05 & 0.80 & 111 & 58 & 80 & 84 & 66 & 86 & \textbf{0.832} & \textbf{0.716} & 0.814 & 0.784 & \textbf{0.739} & 0.785 \\
 &  &  & 0.10 & 0.80 & 81 & 43 & 61 & 61 & 48 & 63 & 0.829 & \textbf{0.718} & 0.812 & 0.781 & \textbf{0.736} & 0.784 \\
 &  &  & 0.05 & 0.90 & 154 & 80 & 120 & 116 & 92 & 120 & 0.929 & \textbf{0.814} & 0.917 & 0.885 & \textbf{0.843} & 0.890 \\
 &  &  & 0.10 & 0.90 & 118 & 62 & 97 & 89 & 70 & 92 & 0.926 & \textbf{0.816} & 0.918 & 0.884 & \textbf{0.840} & 0.888 \\ \hline
\multicolumn{1}{r}{6} & \multicolumn{1}{r}{0.4} & \multicolumn{1}{r}{0.5} & 0.05 & 0.80 & 170 & 136 & 158 & 181 & 165 & 167 & 0.801 & \textbf{0.759} & 0.808 & 0.815 & 0.798 & 0.799 \\
 &  &  & 0.10 & 0.80 & 124 & 100 & 118 & 132 & 120 & 122 & 0.801 & \textbf{0.763} & 0.809 & 0.814 & 0.797 & 0.799 \\
 &  &  & 0.05 & 0.90 & 235 & 189 & 228 & 251 & 228 & 232 & 0.900 & \textbf{0.862} & 0.910 & 0.914 & 0.897 & 0.899 \\
 &  &  & 0.10 & 0.90 & 180 & 145 & 180 & 193 & 175 & 178 & 0.900 & \textbf{0.863} & 0.912 & 0.913 & 0.897 & 0.899 \\ \hline
\multicolumn{1}{r}{6} & \multicolumn{1}{r}{0.7} & \multicolumn{1}{r}{0.8} & 0.05 & 0.80 & 107 & 94 & 115 & 153 & 144 & 125 & 0.777 & \textbf{0.755} & 0.818 & \textbf{0.850} & \textbf{0.838} & 0.809 \\
 &  &  & 0.10 & 0.80 & 78 & 69 & 87 & 112 & 105 & 91 & 0.779 & \textbf{0.761} & 0.823 & \textbf{0.850} & \textbf{0.838} & 0.809 \\
 &  &  & 0.05 & 0.90 & 149 & 130 & 168 & 212 & 200 & 173 & 0.878 & \textbf{0.853} & 0.918 & \textbf{0.939} & \textbf{0.931} & 0.906 \\
 &  &  & 0.10 & 0.90 & 114 & 100 & 134 & 163 & 153 & 132 & 0.879 & \textbf{0.858} & 0.921 & \textbf{0.938} & 0.929 & 0.905 \\ \hline
\multicolumn{17}{l}{\footnotesize Values not within a prescribed range of $1-\beta=0.8$ or $0.9$, plus or minus $0.03$ are highlighted in \textbf{bold}.}\\
\multicolumn{17}{l}{\footnotesize ident.: the identity transformation based on equation (3); log: the log transformation based on equation (3); log (2): the log transformation based on }\\
\multicolumn{17}{l}{\footnotesize equation (2); log-log: the log transformation based on equation (3); logit: the logit transformation based on equation (3); arcsin: the arcsine square-root}\\
\multicolumn{17}{l}{\footnotesize transformation based on equation (3).}
\end{tabular}
\end{table}

\clearpage
\begin{table}[p] 
\centering
\fontsize{13}{13}\selectfont
\caption{Simulation results for power: $\mathrm{Weibull}(\lambda, k=0.5)$ survival distribution and no random censoring were assumed.}
\label{tab:s3}
\begin{tabular}{lllrrrrrrrrrrrrrr} \hline
\multicolumn{1}{c}{\multirow{2}{*}{$b$}} & \multicolumn{1}{c}{\multirow{2}{*}{$S_0(t)$}} & \multicolumn{1}{c}{\multirow{2}{*}{$S_1(t)$}} & \multicolumn{1}{c}{\multirow{2}{*}{$\alpha$}} & \multicolumn{1}{c}{\multirow{2}{*}{$1-\beta$}} & \multicolumn{6}{c}{Sample size} & \multicolumn{6}{c}{Empirical power} \\
 &  &  &  &  & \multicolumn{1}{c}{ident.} & \multicolumn{1}{c}{log} & \multicolumn{1}{c}{log (2)} & \multicolumn{1}{c}{log-log} & \multicolumn{1}{c}{logit} & \multicolumn{1}{c}{arcsin} & \multicolumn{1}{c}{ident.} & \multicolumn{1}{c}{log} & \multicolumn{1}{c}{log (2)} & \multicolumn{1}{c}{log-log} & \multicolumn{1}{c}{logit} & \multicolumn{1}{c}{arcsin} \\ \hline
\multicolumn{1}{r}{12} & \multicolumn{1}{r}{0.1} & \multicolumn{1}{r}{0.2} & 0.05 & 0.80 & 99 & 52 & 71 & 75 & 59 & 77 & \textbf{0.861} & \textbf{0.739} & 0.786 & \textbf{0.761} & \textbf{0.769} & 0.794 \\
 &  &  & 0.10 & 0.80 & 73 & 38 & 54 & 55 & 43 & 56 & 0.817 & \textbf{0.660} & 0.780 & 0.798 & 0.784 & 0.815 \\
 &  &  & 0.05 & 0.90 & 138 & 72 & 106 & 104 & 82 & 107 & \textbf{0.939} & \textbf{0.802} & 0.921 & 0.907 & \textbf{0.861} & 0.884 \\
 &  &  & 0.10 & 0.90 & 106 & 55 & 86 & 80 & 63 & 82 & 0.921 & \textbf{0.798} & 0.901 & 0.900 & \textbf{0.836} & 0.916 \\ \hline
\multicolumn{1}{r}{12} & \multicolumn{1}{r}{0.4} & \multicolumn{1}{r}{0.5} & 0.05 & 0.80 & 155 & 125 & 144 & 166 & 151 & 153 & 0.789 & \textbf{0.762} & 0.820 & 0.803 & 0.792 & 0.791 \\
 &  &  & 0.10 & 0.80 & 113 & 91 & 108 & 121 & 110 & 112 & 0.826 & \textbf{0.735} & 0.806 & 0.818 & 0.804 & 0.802 \\
 &  &  & 0.05 & 0.90 & 215 & 172 & 208 & 229 & 209 & 212 & 0.890 & 0.874 & 0.906 & 0.907 & 0.893 & 0.904 \\
 &  &  & 0.10 & 0.90 & 165 & 132 & 164 & 176 & 160 & 163 & 0.894 & 0.871 & 0.908 & 0.924 & 0.911 & 0.895 \\ \hline
\multicolumn{1}{r}{12} & \multicolumn{1}{r}{0.7} & \multicolumn{1}{r}{0.8} & 0.05 & 0.80 & 99 & 87 & 106 & 142 & 134 & 115 & \textbf{0.755} & \textbf{0.719} & 0.791 & \textbf{0.857} & \textbf{0.845} & 0.795 \\
 &  &  & 0.10 & 0.80 & 73 & 64 & 80 & 103 & 97 & 84 & 0.805 & 0.803 & \textbf{0.837} & \textbf{0.832} & \textbf{0.851} & 0.773 \\
 &  &  & 0.05 & 0.90 & 138 & 121 & 155 & 196 & 185 & 160 & 0.893 & \textbf{0.836} & \textbf{0.931} & \textbf{0.949} & \textbf{0.938} & 0.898 \\
 &  &  & 0.10 & 0.90 & 106 & 93 & 123 & 150 & 142 & 123 & 0.899 & \textbf{0.845} & 0.906 & \textbf{0.934} & 0.929 & 0.906 \\ \hline
\multicolumn{1}{r}{6} & \multicolumn{1}{r}{0.1} & \multicolumn{1}{r}{0.2} & 0.05 & 0.80 & 107 & 56 & 76 & 80 & 64 & 83 & \textbf{0.834} & \textbf{0.717} & 0.811 & 0.782 & \textbf{0.746} & 0.788 \\
 &  &  & 0.10 & 0.80 & 78 & 41 & 59 & 59 & 46 & 61 & \textbf{0.831} & \textbf{0.721} & 0.814 & 0.782 & \textbf{0.736} & 0.790 \\
 &  &  & 0.05 & 0.90 & 147 & 77 & 115 & 111 & 88 & 115 & 0.928 & \textbf{0.815} & 0.918 & 0.885 & \textbf{0.843} & 0.890 \\
 &  &  & 0.10 & 0.90 & 113 & 59 & 93 & 85 & 67 & 88 & 0.927 & \textbf{0.814} & 0.919 & 0.883 & \textbf{0.841} & 0.888 \\ \hline
\multicolumn{1}{r}{6} & \multicolumn{1}{r}{0.4} & \multicolumn{1}{r}{0.5} & 0.05 & 0.80 & 163 & 131 & 152 & 175 & 159 & 161 & 0.800 & \textbf{0.760} & 0.809 & 0.817 & 0.798 & 0.799 \\
 &  &  & 0.10 & 0.80 & 119 & 96 & 114 & 127 & 116 & 118 & 0.800 & \textbf{0.762} & 0.811 & 0.814 & 0.798 & 0.800 \\
 &  &  & 0.05 & 0.90 & 226 & 182 & 220 & 242 & 220 & 223 & 0.900 & \textbf{0.862} & 0.910 & 0.913 & 0.898 & 0.899 \\
 &  &  & 0.10 & 0.90 & 174 & 140 & 173 & 186 & 169 & 171 & 0.901 & \textbf{0.864} & 0.911 & 0.913 & 0.897 & 0.899 \\ \hline
\multicolumn{1}{r}{6} & \multicolumn{1}{r}{0.7} & \multicolumn{1}{r}{0.8} & 0.05 & 0.80 & 104 & 91 & 111 & 148 & 140 & 121 & 0.779 & \textbf{0.757} & 0.817 & \textbf{0.850} & \textbf{0.841} & 0.809 \\
 &  &  & 0.10 & 0.80 & 76 & 67 & 84 & 108 & 102 & 88 & 0.780 & \textbf{0.765} & 0.821 & \textbf{0.850} & \textbf{0.839} & 0.809 \\
 &  &  & 0.05 & 0.90 & 144 & 126 & 163 & 205 & 193 & 167 & 0.878 & \textbf{0.855} & 0.918 & \textbf{0.939} & \textbf{0.930} & 0.906 \\
 &  &  & 0.10 & 0.90 & 110 & 97 & 129 & 157 & 148 & 128 & 0.879 & \textbf{0.860} & 0.920 & \textbf{0.938} & \textbf{0.930} & 0.906 \\ \hline
\multicolumn{17}{l}{\footnotesize Values not within a prescribed range of $1-\beta=0.8$ or $0.9$, plus or minus $0.03$ are highlighted in \textbf{bold}.}\\
\multicolumn{17}{l}{\footnotesize ident.: the identity transformation based on equation (3); log: the log transformation based on equation (3); log (2): the log transformation based on }\\
\multicolumn{17}{l}{\footnotesize equation (2); log-log: the log transformation based on equation (3); logit: the logit transformation based on equation (3); arcsin: the arcsine square-root}\\
\multicolumn{17}{l}{\footnotesize transformation based on equation (3).}
\end{tabular}
\end{table}

\clearpage
\begin{table}[p] 
\centering
\fontsize{13}{13}\selectfont
\caption{Simulation results for power: $\mathrm{Weibull}(\lambda, k=2)$ survival distribution and no random censoring were assumed.}
\label{tab:s4}
\begin{tabular}{lllrrrrrrrrrrrrrr} \hline
\multicolumn{1}{c}{\multirow{2}{*}{$b$}} & \multicolumn{1}{c}{\multirow{2}{*}{$S_0(t)$}} & \multicolumn{1}{c}{\multirow{2}{*}{$S_1(t)$}} & \multicolumn{1}{c}{\multirow{2}{*}{$\alpha$}} & \multicolumn{1}{c}{\multirow{2}{*}{$1-\beta$}} & \multicolumn{6}{c}{Sample size} & \multicolumn{6}{c}{Empirical power} \\
 &  &  &  &  & \multicolumn{1}{c}{ident.} & \multicolumn{1}{c}{log} & \multicolumn{1}{c}{log (2)} & \multicolumn{1}{c}{log-log} & \multicolumn{1}{c}{logit} & \multicolumn{1}{c}{arcsin} & \multicolumn{1}{c}{ident.} & \multicolumn{1}{c}{log} & \multicolumn{1}{c}{log (2)} & \multicolumn{1}{c}{log-log} & \multicolumn{1}{c}{logit} & \multicolumn{1}{c}{arcsin} \\ \hline
\multicolumn{1}{r}{12} & \multicolumn{1}{r}{0.1} & \multicolumn{1}{r}{0.2} & 0.05 & 0.80 & 99 & 52 & 71 & 75 & 59 & 77 & \textbf{0.861} & \textbf{0.739} & 0.786 & \textbf{0.761} & \textbf{0.769} & 0.794 \\
 &  &  & 0.10 & 0.80 & 73 & 38 & 54 & 55 & 43 & 56 & 0.817 & \textbf{0.660} & 0.780 & 0.798 & 0.784 & 0.815 \\
 &  &  & 0.05 & 0.90 & 138 & 72 & 106 & 104 & 82 & 107 & \textbf{0.939} & \textbf{0.802} & 0.921 & 0.907 & \textbf{0.861} & 0.884 \\
 &  &  & 0.10 & 0.90 & 106 & 55 & 86 & 80 & 63 & 82 & 0.921 & \textbf{0.798} & 0.901 & 0.900 & \textbf{0.836} & 0.916 \\ \hline
\multicolumn{1}{r}{12} & \multicolumn{1}{r}{0.4} & \multicolumn{1}{r}{0.5} & 0.05 & 0.80 & 155 & 125 & 144 & 166 & 151 & 153 & 0.789 & \textbf{0.762} & 0.820 & 0.803 & 0.792 & 0.791 \\
 &  &  & 0.10 & 0.80 & 113 & 91 & 108 & 121 & 110 & 112 & 0.826 & \textbf{0.735} & 0.806 & 0.818 & 0.804 & 0.802 \\
 &  &  & 0.05 & 0.90 & 215 & 172 & 208 & 229 & 209 & 212 & 0.890 & 0.874 & 0.906 & 0.907 & 0.893 & 0.904 \\
 &  &  & 0.10 & 0.90 & 165 & 132 & 164 & 176 & 160 & 163 & 0.894 & 0.871 & 0.908 & 0.924 & 0.911 & 0.895 \\ \hline
\multicolumn{1}{r}{12} & \multicolumn{1}{r}{0.7} & \multicolumn{1}{r}{0.8} & 0.05 & 0.80 & 99 & 87 & 106 & 142 & 134 & 115 & \textbf{0.755} & \textbf{0.719} & 0.791 & \textbf{0.857} & \textbf{0.845} & 0.795 \\
 &  &  & 0.10 & 0.80 & 73 & 64 & 80 & 103 & 97 & 84 & 0.805 & 0.803 & \textbf{0.837} & \textbf{0.832} & \textbf{0.851} & 0.773 \\
 &  &  & 0.05 & 0.90 & 138 & 121 & 155 & 196 & 185 & 160 & 0.893 & \textbf{0.836} & \textbf{0.931} & \textbf{0.949} & \textbf{0.938} & 0.898 \\
 &  &  & 0.10 & 0.90 & 106 & 93 & 123 & 150 & 142 & 123 & 0.899 & \textbf{0.845} & 0.906 & \textbf{0.934} & 0.929 & 0.906 \\ \hline
\multicolumn{1}{r}{6} & \multicolumn{1}{r}{0.1} & \multicolumn{1}{r}{0.2} & 0.05 & 0.80 & 117 & 61 & 84 & 88 & 70 & 91 & \textbf{0.833} & \textbf{0.715} & 0.813 & 0.783 & \textbf{0.742} & 0.787 \\
 &  &  & 0.10 & 0.80 & 85 & 45 & 64 & 64 & 51 & 66 & 0.829 & \textbf{0.716} & 0.812 & 0.780 & \textbf{0.739} & 0.783 \\
 &  &  & 0.05 & 0.90 & 162 & 84 & 126 & 122 & 96 & 126 & 0.929 & \textbf{0.813} & 0.917 & 0.885 & \textbf{0.841} & 0.889 \\
 &  &  & 0.10 & 0.90 & 124 & 65 & 102 & 94 & 74 & 96 & 0.926 & \textbf{0.816} & 0.918 & 0.885 & \textbf{0.841} & 0.886 \\ \hline
\multicolumn{1}{r}{6} & \multicolumn{1}{r}{0.4} & \multicolumn{1}{r}{0.5} & 0.05 & 0.80 & 178 & 143 & 166 & 190 & 173 & 176 & 0.800 & \textbf{0.761} & 0.808 & 0.815 & 0.797 & 0.799 \\
 &  &  & 0.10 & 0.80 & 130 & 104 & 124 & 139 & 126 & 128 & 0.801 & \textbf{0.761} & 0.809 & 0.815 & 0.797 & 0.799 \\
 &  &  & 0.05 & 0.90 & 246 & 198 & 240 & 263 & 240 & 243 & 0.900 & \textbf{0.861} & 0.911 & 0.912 & 0.898 & 0.899 \\
 &  &  & 0.10 & 0.90 & 189 & 152 & 189 & 202 & 184 & 187 & 0.900 & \textbf{0.863} & 0.912 & 0.912 & 0.897 & 0.899 \\ \hline
\multicolumn{1}{r}{6} & \multicolumn{1}{r}{0.7} & \multicolumn{1}{r}{0.8} & 0.05 & 0.80 & 113 & 99 & 121 & 161 & 151 & 131 & 0.779 & \textbf{0.757} & 0.819 & \textbf{0.851} & \textbf{0.838} & 0.809 \\
 &  &  & 0.10 & 0.80 & 82 & 72 & 91 & 117 & 110 & 95 & 0.780 & \textbf{0.760} & 0.821 & \textbf{0.849} & \textbf{0.837} & 0.808 \\
 &  &  & 0.05 & 0.90 & 156 & 137 & 176 & 222 & 209 & 181 & 0.878 & \textbf{0.855} & 0.917 & \textbf{0.939} & \textbf{0.930} & 0.905 \\
 &  &  & 0.10 & 0.90 & 120 & 105 & 140 & 171 & 161 & 139 & 0.880 & \textbf{0.857} & 0.920 & \textbf{0.938} & \textbf{0.930} & 0.906 \\ \hline
\multicolumn{17}{l}{\footnotesize Values not within a prescribed range of $1-\beta=0.8$ or $0.9$, plus or minus $0.03$ are highlighted in \textbf{bold}.}\\
\multicolumn{17}{l}{\footnotesize ident.: the identity transformation based on equation (3); log: the log transformation based on equation (3); log (2): the log transformation based on }\\
\multicolumn{17}{l}{\footnotesize equation (2); log-log: the log transformation based on equation (3); logit: the logit transformation based on equation (3); arcsin: the arcsine square-root}\\
\multicolumn{17}{l}{\footnotesize transformation based on equation (3).}
\end{tabular}
\end{table}

\clearpage
\begin{table}[p] 
\centering
\fontsize{13}{13}\selectfont
\caption{Simulation results for power: exponential survival distribution and random censoring were assumed.}
\label{tab:s5}
\begin{tabular}{lllrrrrrrrrrrrrrr} \hline
\multicolumn{1}{c}{\multirow{2}{*}{$b$}} & \multicolumn{1}{c}{\multirow{2}{*}{$S_0(t)$}} & \multicolumn{1}{c}{\multirow{2}{*}{$S_1(t)$}} & \multicolumn{1}{c}{\multirow{2}{*}{$\alpha$}} & \multicolumn{1}{c}{\multirow{2}{*}{$1-\beta$}} & \multicolumn{6}{c}{Sample size} & \multicolumn{6}{c}{Empirical power} \\
 &  &  &  &  & \multicolumn{1}{c}{ident.} & \multicolumn{1}{c}{log} & \multicolumn{1}{c}{log (2)} & \multicolumn{1}{c}{log-log} & \multicolumn{1}{c}{logit} & \multicolumn{1}{c}{arcsin} & \multicolumn{1}{c}{ident.} & \multicolumn{1}{c}{log} & \multicolumn{1}{c}{log (2)} & \multicolumn{1}{c}{log-log} & \multicolumn{1}{c}{logit} & \multicolumn{1}{c}{arcsin} \\ \hline
\multicolumn{1}{r}{12} & \multicolumn{1}{r}{0.1} & \multicolumn{1}{r}{0.2} & 0.05 & 0.80 & 129 & 67 & 98 & 97 & 77 & 100 & \textbf{0.832} & \textbf{0.713} & 0.829 & 0.782 & \textbf{0.740} & 0.785 \\
 &  &  & 0.10 & 0.80 & 94 & 49 & 75 & 71 & 56 & 73 & 0.829 & \textbf{0.713} & 0.827 & 0.780 & \textbf{0.738} & 0.783 \\
 &  &  & 0.05 & 0.90 & 178 & 93 & 149 & 134 & 106 & 138 & 0.928 & \textbf{0.814} & \textbf{0.930} & 0.884 & \textbf{0.840} & 0.887 \\
 &  &  & 0.10 & 0.90 & 137 & 71 & 121 & 103 & 81 & 106 & 0.926 & \textbf{0.813} & \textbf{0.930} & 0.882 & \textbf{0.839} & 0.885 \\ \hline
\multicolumn{1}{r}{12} & \multicolumn{1}{r}{0.4} & \multicolumn{1}{r}{0.5} & 0.05 & 0.80 & 171 & 137 & 161 & 183 & 166 & 169 & 0.802 & \textbf{0.761} & 0.813 & 0.818 & 0.798 & 0.801 \\
 &  &  & 0.10 & 0.80 & 125 & 100 & 121 & 133 & 121 & 123 & 0.802 & \textbf{0.762} & 0.815 & 0.815 & 0.798 & 0.800 \\
 &  &  & 0.05 & 0.90 & 237 & 190 & 233 & 253 & 230 & 233 & 0.901 & \textbf{0.863} & 0.914 & 0.915 & 0.898 & 0.899 \\
 &  &  & 0.10 & 0.90 & 182 & 146 & 184 & 194 & 177 & 179 & 0.902 & \textbf{0.865} & 0.916 & 0.914 & 0.899 & 0.900 \\ \hline
\multicolumn{1}{r}{12} & \multicolumn{1}{r}{0.7} & \multicolumn{1}{r}{0.8} & 0.05 & 0.80 & 102 & 90 & 110 & 146 & 137 & 119 & 0.779 & \textbf{0.762} & 0.821 & \textbf{0.851} & \textbf{0.839} & 0.811 \\
 &  &  & 0.10 & 0.80 & 75 & 66 & 83 & 106 & 100 & 87 & 0.787 & \textbf{0.769} & 0.822 & \textbf{0.849} & \textbf{0.839} & 0.814 \\
 &  &  & 0.05 & 0.90 & 142 & 124 & 161 & 202 & 190 & 164 & 0.880 & \textbf{0.855} & 0.919 & \textbf{0.940} & \textbf{0.931} & 0.905 \\
 &  &  & 0.10 & 0.90 & 109 & 95 & 128 & 155 & 146 & 126 & 0.882 & \textbf{0.856} & 0.922 & \textbf{0.939} & \textbf{0.930} & 0.906 \\ \hline 
\multicolumn{1}{r}{6} & \multicolumn{1}{r}{0.1} & \multicolumn{1}{r}{0.2} & 0.05 & 0.80 & 145 & 76 & 111 & 109 & 87 & 113 & \textbf{0.861} & \textbf{0.747} & \textbf{0.856} & 0.812 & 0.773 & 0.817 \\
 &  &  & 0.10 & 0.80 & 106 & 55 & 86 & 80 & 63 & 82 & \textbf{0.854} & \textbf{0.739} & \textbf{0.854} & 0.807 & \textbf{0.764} & 0.809 \\
 &  &  & 0.05 & 0.90 & 201 & 105 & 170 & 151 & 120 & 156 & \textbf{0.946} & \textbf{0.841} & \textbf{0.948} & 0.906 & \textbf{0.867} & 0.910 \\
 &  &  & 0.10 & 0.90 & 154 & 81 & 139 & 116 & 92 & 120 & \textbf{0.941} & \textbf{0.839} & \textbf{0.947} & 0.902 & \textbf{0.863} & 0.906 \\ \hline
\multicolumn{1}{r}{6} & \multicolumn{1}{r}{0.4} & \multicolumn{1}{r}{0.5} & 0.05 & 0.80 & 188 & 151 & 178 & 201 & 183 & 185 & \textbf{0.841} & 0.801 & \textbf{0.852} & \textbf{0.855} & \textbf{0.839} & \textbf{0.839} \\
 &  &  & 0.10 & 0.80 & 137 & 110 & 133 & 146 & 133 & 135 & \textbf{0.835} & 0.796 & \textbf{0.847} & \textbf{0.849} & \textbf{0.831} & \textbf{0.833} \\
 &  &  & 0.05 & 0.90 & 260 & 209 & 257 & 278 & 253 & 256 & 0.928 & 0.894 & \textbf{0.938} & \textbf{0.939} & 0.926 & 0.927 \\
 &  &  & 0.10 & 0.90 & 199 & 160 & 203 & 213 & 194 & 197 & 0.925 & 0.891 & \textbf{0.937} & \textbf{0.936} & 0.923 & 0.924 \\ \hline
\multicolumn{1}{r}{6} & \multicolumn{1}{r}{0.7} & \multicolumn{1}{r}{0.8} & 0.05 & 0.80 & 111 & 97 & 119 & 158 & 149 & 129 & 0.804 & 0.781 & \textbf{0.843} & \textbf{0.874} & \textbf{0.864} & \textbf{0.835} \\
 &  &  & 0.10 & 0.80 & 81 & 71 & 90 & 115 & 109 & 94 & 0.804 & 0.782 & \textbf{0.843} & \textbf{0.869} & \textbf{0.860} & \textbf{0.832} \\
 &  &  & 0.05 & 0.90 & 153 & 134 & 175 & 218 & 206 & 178 & 0.897 & 0.875 & \textbf{0.935} & \textbf{0.952} & \textbf{0.945} & 0.923 \\
 &  &  & 0.10 & 0.90 & 118 & 103 & 139 & 168 & 158 & 137 & 0.898 & 0.876 & \textbf{0.936} & \textbf{0.951} & \textbf{0.943} & 0.922 \\ \hline
\multicolumn{17}{l}{\footnotesize Values not within a prescribed range of $1-\beta=0.8$ or $0.9$, plus or minus $0.03$ are highlighted in \textbf{bold}.}\\
\multicolumn{17}{l}{\footnotesize ident.: the identity transformation based on equation (3); log: the log transformation based on equation (3); log (2): the log transformation based on }\\
\multicolumn{17}{l}{\footnotesize equation (2); log-log: the log transformation based on equation (3); logit: the logit transformation based on equation (3); arcsin: the arcsine square-root}\\
\multicolumn{17}{l}{\footnotesize transformation based on equation (3).}
\end{tabular}
\end{table}

\clearpage
\begin{table}[p] 
\centering
\fontsize{13}{13}\selectfont
\caption{Simulation results for power: $\mathrm{Weibull}(\lambda, k=0.5)$ survival distribution and random censoring were assumed.}
\label{tab:s6}
\begin{tabular}{lllrrrrrrrrrrrrrr} \hline
\multicolumn{1}{c}{\multirow{2}{*}{$b$}} & \multicolumn{1}{c}{\multirow{2}{*}{$S_0(t)$}} & \multicolumn{1}{c}{\multirow{2}{*}{$S_1(t)$}} & \multicolumn{1}{c}{\multirow{2}{*}{$\alpha$}} & \multicolumn{1}{c}{\multirow{2}{*}{$1-\beta$}} & \multicolumn{6}{c}{Sample size} & \multicolumn{6}{c}{Empirical power} \\
 &  &  &  &  & \multicolumn{1}{c}{ident.} & \multicolumn{1}{c}{log} & \multicolumn{1}{c}{log (2)} & \multicolumn{1}{c}{log-log} & \multicolumn{1}{c}{logit} & \multicolumn{1}{c}{arcsin} & \multicolumn{1}{c}{ident.} & \multicolumn{1}{c}{log} & \multicolumn{1}{c}{log (2)} & \multicolumn{1}{c}{log-log} & \multicolumn{1}{c}{logit} & \multicolumn{1}{c}{arcsin} \\ \hline
\multicolumn{1}{r}{12} & \multicolumn{1}{r}{0.1} & \multicolumn{1}{r}{0.2} & 0.05 & 0.80 & 129 & 67 & 98 & 97 & 77 & 100 & \textbf{0.832} & \textbf{0.713} & 0.829 & 0.782 & \textbf{0.740} & 0.785 \\
 &  &  & 0.10 & 0.80 & 94 & 49 & 75 & 71 & 56 & 73 & 0.829 & \textbf{0.713} & 0.827 & 0.780 & \textbf{0.738} & 0.783 \\
 &  &  & 0.05 & 0.90 & 178 & 93 & 149 & 134 & 106 & 138 & 0.928 & \textbf{0.814} & \textbf{0.930} & 0.884 & \textbf{0.840} & 0.887 \\
 &  &  & 0.10 & 0.90 & 137 & 71 & 121 & 103 & 81 & 106 & 0.926 & \textbf{0.813} & \textbf{0.930} & 0.882 & \textbf{0.839} & 0.885 \\ \hline
\multicolumn{1}{r}{12} & \multicolumn{1}{r}{0.4} & \multicolumn{1}{r}{0.5} & 0.05 & 0.80 & 171 & 137 & 161 & 183 & 166 & 169 & 0.802 & \textbf{0.761} & 0.813 & 0.818 & 0.798 & 0.801 \\
 &  &  & 0.10 & 0.80 & 125 & 100 & 121 & 133 & 121 & 123 & 0.802 & \textbf{0.762} & 0.815 & 0.815 & 0.798 & 0.800 \\
 &  &  & 0.05 & 0.90 & 237 & 190 & 233 & 253 & 230 & 233 & 0.901 & \textbf{0.863} & 0.914 & 0.915 & 0.898 & 0.899 \\
 &  &  & 0.10 & 0.90 & 182 & 146 & 184 & 194 & 177 & 179 & 0.902 & \textbf{0.865} & 0.916 & 0.914 & 0.899 & 0.900 \\ \hline
\multicolumn{1}{r}{12} & \multicolumn{1}{r}{0.7} & \multicolumn{1}{r}{0.8} & 0.05 & 0.80 & 102 & 90 & 110 & 146 & 137 & 119 & 0.779 & \textbf{0.762} & 0.822 & \textbf{0.851} & \textbf{0.839} & 0.811 \\
 &  &  & 0.10 & 0.80 & 75 & 66 & 83 & 106 & 100 & 87 & 0.787 & \textbf{0.769} & 0.822 & \textbf{0.849} & \textbf{0.839} & 0.814 \\
 &  &  & 0.05 & 0.90 & 142 & 124 & 161 & 202 & 190 & 164 & 0.880 & \textbf{0.855} & 0.919 & \textbf{0.940} & \textbf{0.931} & 0.905 \\
 &  &  & 0.10 & 0.90 & 109 & 95 & 128 & 155 & 146 & 126 & 0.882 & \textbf{0.856} & 0.922 & \textbf{0.939} & \textbf{0.930} & 0.906 \\ \hline
\multicolumn{1}{r}{6} & \multicolumn{1}{r}{0.1} & \multicolumn{1}{r}{0.2} & 0.05 & 0.80 & 139 & 73 & 106 & 105 & 83 & 108 & \textbf{0.848} & \textbf{0.735} & \textbf{0.845} & 0.802 & \textbf{0.759} & 0.804 \\
 &  &  & 0.10 & 0.80 & 101 & 53 & 82 & 76 & 60 & 79 & \textbf{0.842} & \textbf{0.731} & \textbf{0.843} & 0.794 & \textbf{0.752} & 0.799 \\
 &  &  & 0.05 & 0.90 & 192 & 100 & 162 & 145 & 114 & 149 & \textbf{0.938} & \textbf{0.828} & \textbf{0.941} & 0.898 & \textbf{0.854} & 0.900 \\
 &  &  & 0.10 & 0.90 & 148 & 77 & 132 & 111 & 88 & 115 & \textbf{0.936} & \textbf{0.827} & \textbf{0.940} & 0.893 & \textbf{0.853} & 0.898 \\ \hline
\multicolumn{1}{r}{6} & \multicolumn{1}{r}{0.4} & \multicolumn{1}{r}{0.5} & 0.05 & 0.80 & 181 & 145 & 171 & 193 & 176 & 178 & 0.825 & 0.784 & \textbf{0.835} & \textbf{0.839} & 0.822 & 0.823 \\
 &  &  & 0.10 & 0.80 & 132 & 106 & 128 & 141 & 128 & 130 & 0.822 & 0.782 & \textbf{0.833} & \textbf{0.836} & 0.817 & 0.819 \\
 &  &  & 0.05 & 0.90 & 250 & 201 & 247 & 267 & 243 & 247 & 0.917 & 0.881 & 0.928 & 0.929 & 0.914 & 0.916 \\
 &  &  & 0.10 & 0.90 & 192 & 154 & 195 & 205 & 187 & 189 & 0.915 & 0.880 & 0.928 & 0.926 & 0.913 & 0.914 \\ \hline
\multicolumn{1}{r}{6} & \multicolumn{1}{r}{0.7} & \multicolumn{1}{r}{0.8} & 0.05 & 0.80 & 107 & 94 & 115 & 153 & 144 & 124 & 0.793 & 0.772 & \textbf{0.833} & \textbf{0.865} & \textbf{0.854} & 0.822 \\
 &  &  & 0.10 & 0.80 & 78 & 69 & 87 & 111 & 105 & 91 & 0.792 & 0.773 & \textbf{0.835} & \textbf{0.860} & \textbf{0.851} & 0.823 \\
 &  &  & 0.05 & 0.90 & 148 & 130 & 169 & 211 & 199 & 172 & 0.889 & \textbf{0.867} & 0.929 & \textbf{0.947} & \textbf{0.940} & 0.916 \\
 &  &  & 0.10 & 0.90 & 114 & 100 & 134 & 162 & 153 & 132 & 0.890 & \textbf{0.869} & 0.930 & \textbf{0.945} & \textbf{0.938} & 0.915 \\ \hline
\multicolumn{17}{l}{\footnotesize Values not within a prescribed range of $1-\beta=0.8$ or $0.9$, plus or minus $0.03$ are highlighted in \textbf{bold}.}\\
\multicolumn{17}{l}{\footnotesize ident.: the identity transformation based on equation (3); log: the log transformation based on equation (3); log (2): the log transformation based on }\\
\multicolumn{17}{l}{\footnotesize equation (2); log-log: the log transformation based on equation (3); logit: the logit transformation based on equation (3); arcsin: the arcsine square-root}\\
\multicolumn{17}{l}{\footnotesize transformation based on equation (3).}
\end{tabular}
\end{table}

\clearpage
\begin{table}[p] 
\centering
\fontsize{13}{13}\selectfont
\caption{Simulation results for power: $\mathrm{Weibull}(\lambda, k=2)$ survival distribution and random censoring were assumed.}
\label{tab:s7}
\begin{tabular}{lllrrrrrrrrrrrrrr} \hline
\multicolumn{1}{c}{\multirow{2}{*}{$b$}} & \multicolumn{1}{c}{\multirow{2}{*}{$S_0(t)$}} & \multicolumn{1}{c}{\multirow{2}{*}{$S_1(t)$}} & \multicolumn{1}{c}{\multirow{2}{*}{$\alpha$}} & \multicolumn{1}{c}{\multirow{2}{*}{$1-\beta$}} & \multicolumn{6}{c}{Sample size} & \multicolumn{6}{c}{Empirical power} \\
 &  &  &  &  & \multicolumn{1}{c}{ident.} & \multicolumn{1}{c}{log} & \multicolumn{1}{c}{log (2)} & \multicolumn{1}{c}{log-log} & \multicolumn{1}{c}{logit} & \multicolumn{1}{c}{arcsin} & \multicolumn{1}{c}{ident.} & \multicolumn{1}{c}{log} & \multicolumn{1}{c}{log (2)} & \multicolumn{1}{c}{log-log} & \multicolumn{1}{c}{logit} & \multicolumn{1}{c}{arcsin} \\ \hline
\multicolumn{1}{r}{12} & \multicolumn{1}{r}{0.1} & \multicolumn{1}{r}{0.2} & 0.05 & 0.80 & 129 & 67 & 98 & 97 & 77 & 100 & \textbf{0.832} & \textbf{0.713} & 0.829 & 0.782 & \textbf{0.740} & 0.785 \\
 &  &  & 0.10 & 0.80 & 94 & 49 & 75 & 71 & 56 & 73 & 0.829 & \textbf{0.713} & 0.827 & 0.780 & \textbf{0.738} & 0.783 \\
 &  &  & 0.05 & 0.90 & 178 & 93 & 149 & 134 & 106 & 138 & 0.928 & \textbf{0.814} & \textbf{0.930} & 0.884 & \textbf{0.840} & 0.887 \\
 &  &  & 0.10 & 0.90 & 137 & 71 & 121 & 103 & 81 & 106 & 0.926 & \textbf{0.813} & \textbf{0.930} & 0.882 & \textbf{0.839} & 0.885 \\ \hline
\multicolumn{1}{r}{12} & \multicolumn{1}{r}{0.4} & \multicolumn{1}{r}{0.5} & 0.05 & 0.80 & 171 & 137 & 161 & 183 & 166 & 169 & 0.802 & \textbf{0.761} & 0.813 & 0.818 & 0.798 & 0.801 \\
 &  &  & 0.10 & 0.80 & 125 & 100 & 121 & 133 & 121 & 123 & 0.802 & \textbf{0.762} & 0.815 & 0.815 & 0.798 & 0.800 \\
 &  &  & 0.05 & 0.90 & 237 & 190 & 233 & 253 & 230 & 233 & 0.901 & \textbf{0.863} & 0.914 & 0.915 & 0.898 & 0.899 \\
 &  &  & 0.10 & 0.90 & 182 & 146 & 184 & 194 & 177 & 179 & 0.902 & \textbf{0.865} & 0.916 & 0.914 & 0.899 & 0.900 \\ \hline
\multicolumn{1}{r}{12} & \multicolumn{1}{r}{0.7} & \multicolumn{1}{r}{0.8} & 0.05 & 0.80 & 102 & 90 & 110 & 146 & 137 & 119 & 0.779 & \textbf{0.762} & 0.822 & \textbf{0.851} & \textbf{0.839} & 0.811 \\
 &  &  & 0.10 & 0.80 & 75 & 66 & 83 & 106 & 100 & 87 & 0.786 & \textbf{0.769} & 0.822 & \textbf{0.849} & \textbf{0.839} & 0.814 \\
 &  &  & 0.05 & 0.90 & 142 & 124 & 161 & 202 & 190 & 164 & 0.880 & \textbf{0.855} & 0.919 & \textbf{0.940} & \textbf{0.931} & 0.905 \\
 &  &  & 0.10 & 0.90 & 109 & 95 & 128 & 155 & 146 & 126 & 0.882 & \textbf{0.856} & 0.922 & \textbf{0.939} & \textbf{0.930} & 0.906 \\ \hline
\multicolumn{1}{r}{6} & \multicolumn{1}{r}{0.1} & \multicolumn{1}{r}{0.2} & 0.05 & 0.80 & 153 & 80 & 117 & 115 & 91 & 119 & \textbf{0.879} & \textbf{0.765} & \textbf{0.871} & \textbf{0.831} & 0.789 & \textbf{0.836} \\
 &  &  & 0.10 & 0.80 & 111 & 58 & 90 & 84 & 66 & 87 & \textbf{0.868} & \textbf{0.756} & \textbf{0.866} & 0.822 & 0.779 & 0.827 \\
 &  &  & 0.05 & 0.90 & 211 & 110 & 178 & 159 & 126 & 164 & \textbf{0.955} & \textbf{0.856} & \textbf{0.956} & 0.919 & 0.882 & 0.923 \\
 &  &  & 0.10 & 0.90 & 162 & 85 & 145 & 122 & 97 & 126 & \textbf{0.951} & \textbf{0.853} & \textbf{0.954} & 0.913 & 0.877 & 0.917 \\ \hline
\multicolumn{1}{r}{6} & \multicolumn{1}{r}{0.4} & \multicolumn{1}{r}{0.5} & 0.05 & 0.80 & 197 & 159 & 187 & 211 & 192 & 195 & \textbf{0.863} & 0.825 & \textbf{0.872} & \textbf{0.877} & \textbf{0.860} & \textbf{0.862} \\
 &  &  & 0.10 & 0.80 & 144 & 116 & 140 & 154 & 140 & 142 & \textbf{0.855} & 0.818 & \textbf{0.866} & \textbf{0.868} & \textbf{0.852} & \textbf{0.854} \\
 &  &  & 0.05 & 0.90 & 273 & 219 & 270 & 292 & 266 & 269 & \textbf{0.943} & 0.911 & \textbf{0.951} & \textbf{0.952} & \textbf{0.940} & \textbf{0.941} \\
 &  &  & 0.10 & 0.90 & 210 & 168 & 213 & 224 & 204 & 207 & \textbf{0.939} & 0.906 & \textbf{0.949} & \textbf{0.948} & \textbf{0.936} & \textbf{0.937} \\ \hline
\multicolumn{1}{r}{6} & \multicolumn{1}{r}{0.7} & \multicolumn{1}{r}{0.8} & 0.05 & 0.80 & 116 & 102 & 125 & 165 & 156 & 135 & 0.818 & 0.797 & \textbf{0.856} & \textbf{0.886} & \textbf{0.877} & \textbf{0.849} \\
 &  &  & 0.10 & 0.80 & 85 & 74 & 95 & 121 & 114 & 98 & 0.817 & 0.793 & \textbf{0.856} & \textbf{0.882} & \textbf{0.872} & \textbf{0.842} \\
 &  &  & 0.05 & 0.90 & 161 & 141 & 184 & 229 & 216 & 187 & 0.909 & 0.888 & \textbf{0.944} & \textbf{0.960} & \textbf{0.953} & \textbf{0.934} \\
 &  &  & 0.10 & 0.90 & 123 & 108 & 146 & 176 & 166 & 143 & 0.907 & 0.886 & \textbf{0.944} & \textbf{0.957} & \textbf{0.951} & 0.930 \\ \hline
\multicolumn{17}{l}{\footnotesize Values not within a prescribed range of $1-\beta=0.8$ or $0.9$, plus or minus $0.03$ are highlighted in \textbf{bold}.}\\
\multicolumn{17}{l}{\footnotesize ident.: the identity transformation based on equation (3); log: the log transformation based on equation (3); log (2): the log transformation based on }\\
\multicolumn{17}{l}{\footnotesize equation (2); log-log: the log transformation based on equation (3); logit: the logit transformation based on equation (3); arcsin: the arcsine square-root}\\
\multicolumn{17}{l}{\footnotesize transformation based on equation (3).}
\end{tabular}
\end{table}

\end{landscape}

\end{document}